\documentclass[a4paper,english,aps,twocolumn,floatfix,amsfonts,amssymb,superscriptaddress]{revtex4-1}
\usepackage{graphicx}
\usepackage{epstopdf} 
\usepackage{dcolumn}
\usepackage{bm}
\usepackage{bbm,bbold}
\usepackage{color}
\usepackage{xcolor, soul}
\sethlcolor{green}
\usepackage{amsfonts}
\usepackage{amsmath}
\usepackage{mdframed}
\usepackage[normalem]{ulem}
\usepackage{mathrsfs}   
\usepackage[percent]{overpic}
\usepackage[colorlinks=true,citecolor=blue]{hyperref}
\hypersetup{colorlinks=true,citecolor=blue,linkcolor=blue,urlcolor=blue}
\DeclareRobustCommand{\rchi}{{\mathpalette\irchi\relax}}
\newcommand{\irchi}[2]{\raisebox{\depth}{$#1\chi$}} 

\begin{document}
\title{Many-Body Effects in Third Harmonic Generation of Graphene} 
\author{Habib Rostami}
\email{habib.rostami@su.se}
\affiliation{Nordita and Stockholm University, Roslagstullsbacken 23, SE-106 91 Stockholm, Sweden}
\author{Emmanuele Cappelluti} 
\affiliation{Istituto di Struttura della Materia, CNR (ISM-CNR), 34149 Trieste, Italy} 
\begin{abstract}
The low-energy (intraband) range of the third harmonic generation of graphene in the terahertz regime is governed by the damping terms induced by the interactions. A controlled many-body description of the scattering processes is thus a compelling and desirable requirement. In this paper, using a Kadanoff-Baym approach, we systematically investigate the impact of many-body interaction on the third-harmonic generation (THG) of graphene, taking elastic impurity scattering as a benchmark example. We predict the onset in the mixed inter-intraband regime of novel incoherent features driven by the interaction at four- and five-photon transition frequencies in the third-harmonic optical conductivity with a spectral weight proportional to the scattering rate.We show also that, in spite of the complex many-body physics, the purely intraband term governing the limit $\omega \to 0$ resembles the constraints of the phenomenological model. We ascribe this agreement to the fulfilling of the conservation laws enforced by the conserving approach. The overlap with novel incoherent features and the impact of many-body driven multi-photon vertex couplings limit however severely the validity of phenomenological description.
\end{abstract}
\maketitle
\section{Introduction} 
A simple dimensional analysis shows
that the zero-temperature third-harmonic response of clean graphene scales as $\sigma^{(3)} \sim 1/(\omega^3 |\mu|)$ \cite{Cheng_2014_njp,Cheng_prb_2015,Mikhailov_prb_2016,Rostami_prb_2016,Rostami_prr_2020} where $\mu$ is the chemical potential and $\omega$ is the frequency of the incident field. Comparing to the scaling of linear conductivity at low frequency, $\sigma^{(1)} \sim |\mu|/\omega$, we can thus expect a huge enhancement of the nonlinear optical response at low frequencies. Accordingly, there is surge of experimental interest in exploring nonlinear optics of graphene and other two-dimensional materials in the terahertz frequency range. The effective three-dimensional THG susceptibility of graphene with an effective thickness $\sim 0.1$ nm has been measured as
$ 10^{-19}-10^{-16}$ ${\rm m^2/V^2}$\cite{Kumar_prb_2013,Woodward_2DM_2016,Soavi_natnano_2018,Soavi_acsphotonics_2019} in the near-infrared/visible frequency range $\hbar\omega \sim 100-500 $THz, whereas in the terahertz range ($\hbar\omega\sim 1$THz)
 a susceptibility as large as $10^{-9}$ ${\rm m^2/V^2}$  is obtained \cite{Hafez_nature_2018,Hafez_review_thz}. 
In these works however the nonlinear response was not related to multi-photon absorption/emission but rather to the
underlying nonlinear dependence of the low-frequency intraband linear conductivity
on the electronic temperature ruled by the magnitude of the incident laser power\cite{Mics_nc_2015,Soavi_acsphotonics_2019,Mikhailov_prb_2019}.  
The intrinsic role of multi-photon processes is thus completely missing in the interpretation of this experimental observation.

In common graphene, the terahertz frequency range, 1THz $\sim$ 4meV, is much below the inter-band transition, $\hbar\omega \ll 2|\mu|$, and therefore the optical response is mainly governed by intra-band processes. Even in low-doping regime, the THz-range inter-band transitions will be essentially Pauli-blocked owing to temperature effects, charge-puddle formation and many-body induced band-broadening. The intra-band response is very sensitive to the scattering processes which result in the momentum relaxation of quasiparticles. Therefore, the many-body interaction is expected to have an unavoidable impact on the intra-band optical response of the Dirac fermions in graphene. A compelling study of nonlinear responses is usually a delicate and cumbersome task. Therefore, the number of studies exploring many-body effects on the nonlinear optical response in a systematic way is quite
limited.\cite{Katsnelson_2010,Rostami_prb_2017a,Rostami_prr_2020,Avdoshkin_prl_2020,Du_arXiv_2020,Rostami_Cappeluti_arxiv_2020}.
A practical short-cut for including the impact of the scattering in the analysis is through a phenomenological relaxation rate $\Gamma$ independent from energy and field \cite{Cheng_prb_2015,Mikhailov_prb_2016}. This rough approximation may work qualitatively well in the
high-doping and high-frequency ($|\mu|, |\hbar \omega | \gg \Gamma$)  regime.
However, even in high quality graphene samples
a scattering rate $\Gamma$ not less than $\Gamma\approx 2-5$ THz was estimated
in a wide range $\mu\sim 0-200$ meV \cite{Tan_prl_2007,Horng_prb_2011}.
Therefore, in the terahertz range ($\hbar\omega \sim 1$ THz) we are rather in the regime 
$\hbar \omega < \Gamma$, questioning the validity of a constant-$\Gamma$ model .
Furthermore, in low-frequency regime vertex corrections to the linear ones are expected
to be extremely relevant due 
to the proximity to the multi-photon self-generation \cite{Rostami_Cappeluti_arxiv_2020}.

Aim of the present work is to provide a compelling many-body approach for the nonlinear response of graphene and other two-dimensional
Dirac systems in the terahertz frequency range, where the role of scattering is fundamentally important.
We derive a conserving quantum theory based on the Kadanoff-Baym framework \cite{kadanoff_Baym} which contains both the self-energy and vertex correction effects on the same footing. Our theory is therefore consistent with conservation laws and the gauge invariance. To further preserve the gauge invariance and not violate the Ward's identity, we employ a dimensional regularisation \cite{Leibbrandt_rmp_1975,Peskin} in the evaluation of momentum integrals in Dirac model of graphene for which a high-energy cut-off is unavoidable. For the sake of simplicity, we only focus on the impact of disorder scattering within the self-consistent Born approximation (SCBA). However, our formal and technical development is valid also for scattering with phonons and other electrons in the system. Most of our qualitative predictions
are thus expected to be valid also for other types of scattering processes. 

Our main results can be summarized as: ($i$) we predict the onset of
four- and five-photon transitions in the third harmonic response of graphene.
As a result of such four- and five-photon transitions, the THG of graphene is strongly enhanced in the intra-band regime where
$\hbar\omega$ is smaller than three-photon transition edge.
These novel transitions are intrinsically driven by the interaction with a spectral weight that scales
with the magnitude of the one-particle scattering rate, revealing the intrinsic incoherent character;
($ii$) a strong impact of vertex corrections is revealed owing the presence of many-body-induced two- and three-photon current vertices which are absent in non-interacting Dirac fermions in graphene. A crucial role in this regards is played in particular by the
occurring of the two-photon vertex self-generation in the intra-band terahertz regime, close to the dc limit;
($iii$) in the extreme low-frequency limit, we find a good agreement of the pure intraband term in the presence
of many-body effects with the phenomenological models, as resulting of enforcing a conserving approach.
Departures from this modelling are however observed also at relatively small scattering due
to the onset of incoherent four and five-photon transitions and the two-photon vertex self-generation.
Our theoretical modeling can be generalized in a straightforward way to explain intra-band THG in other two-dimensional materials such as 
transition-metal dichalcogenides, homo- and hetero-bilayer systems, etc.  

The paper is structured in five Sections. In Section \ref{sec:model_method}, we introduce the conserving Baym-Kadanoff
derivation employed to evaluate the nonlinear current within the Dirac model, and we introduce
the elastic impurity-induced self-energy. In Section \ref{sec:KB}, we formally derive the many-body-induced multi-photon vertices based on self-consistent Bethe-Salpeter equations within the Kadanoff-Baym method.  
In Section \ref{sec:Pfunction} we provide all of the analytical relations for the third-order response function using a diagrammatic quantum theory for THG in graphene and for generic two-dimensional Dirac systems. In Section \ref{sec:result_discussion}, we present our numerical results for the real and imaginary parts of the third-harmonic conductivity in graphene and we discuss the onset of novel incoherent-transition peaks, the impact of vertex renormalization and the spectral features of pure intraband processes.. Eventually in Section \ref{sec:summery_conclusion} we provide a summary and conclusion.

\section{Model and Method}\label{sec:model_method}

We use the Dirac Hamiltonian of low-energy carriers in graphene as
\begin{eqnarray}
\hat {\cal H}_{\bf k} 
&=&
 \hbar v  \hat{\bm \sigma}\cdot {\bf k}-\mu_0\hat I ,
\end{eqnarray}
where $v\sim10^6~{\rm m}/{\rm s}$ is the Fermi velocity and where
the Hamiltonian includes the dependence both on the pseudospin sublattice degree of freedom
and on the valley. More explicitly we write
$\hat{\bm \sigma} = (\tau\hat\sigma_x,\hat\sigma_y)$,
where $\hat\sigma_x$, $\hat\sigma_y$ stand for the Pauli matrices in the sublattice basis and $\tau=\pm$
accounts for the two inequivalent valleys in the Brillouin zone of graphene.

In dipole approximation we can model light-matter interaction by applying minimal coupling transformation 
$\hbar{\bf k} \to \hbar{\bf k} + e{\bf A}(t)$ where ${\bf A}(t)$ stands for an external vector potential and
 the corresponding electric field is given by ${\bf E}(t)=-\partial_t {\bf A}(t)$.
For sake of shortness that the speed of light is set $c=1$. 
The inverse of the bare Green's function in the presence of the external
vector potential reads thus:
\begin{eqnarray}
  \hat G^{-1}_0(1,1';{\bf A}) 
  &=&
  \left\{
  i\partial_{t_1}- v   \hat{\bm \sigma}\cdot [-i\hbar {\bm \nabla}_1 + e{\bf A}(1)]+\mu_0
  \right\} 
  \nonumber\\
  &&
  \times
   \delta(1-1')~,
\end{eqnarray} 
where we use the shorthand notation $1\equiv({\bf r}_1,t_1)$ for the space-time coordinate.
In the presence of many-body scattering, it is useful
to introduce an interacting Green's function:
\begin{align}
 \hat G(1,1';{\bf A}) &=-i \langle {\cal T}[ \hat \psi_{\cal H}(1)\hat \psi^\dagger_{\cal H}(1') ] \rangle~,
\end{align}
where $\langle \dots \rangle$ stands for the thermodynamical average, $ {\cal T}$ for the time-ordering operation,
and $\hat \psi_{\cal H}({\bf r},t) $ denotes the field operator  in the Heisenberg picture in the basis of full Hamiltonian ${\cal H}$
which contains kinetics, light-matter and many-body interaction terms.

Using a standard quantum-field formalism,
the effects of the many-body interacting can be conveniently
cast in terms of the many-body self-energy $\hat \Sigma(1,2)$.
Using Dyson recursive relation, the full field-dependent and interacting Green's function is given in terms of a field-dependent self-energy, $\hat \Sigma$, and of a bare Green's function, $\hat G_0$, as follows 
\begin{align}
   \hat G(1,1';{\bf A}) &=   \hat G_0(1,1';{\bf A})   
   \nonumber\\ &
   + \int_{\bar2,\bar3}  \hat G_0(1,\bar 2;{\bf A})   \hat \Sigma(\bar 2,\bar 3;{\bf A})    \hat G(\bar3,1';{\bf A})~. 
\end{align}
Equivalently we have
\begin{align}\label{eq:dyson}
  \hat G^{-1}(1,1';{\bf A}) =  \hat G^{-1}_0(1,1';{\bf A})-  \hat\Sigma(1,1';{\bf A})~.
\end{align}

From now on we assume an external gauge field along y axis, ${\bf A}(1) = A(1) \hat{\bf y}$. 
The thermodynamical physical current, ${\bf J}(1;{\bf A})= J(1;{\bf A}) \hat {\bf y}$, in Dirac systems
can be now obtained as
\begin{align}\label{eq:physical_current}
J(1;{\bf A}) = 
-i \int_{1',1''} {\rm tr} \left[  \hat \Lambda^{(0)}_{1} (1,1';1'')    \hat  G(1,1'^+;{\bf A})  \right]~,
\end{align}
where we denoted $1'^+ \equiv ({\bf r}_{1'}, t_{1'}+0^+)$ and ``${\rm tr}$'' stands for the ``trace'' operation over all spinor indexes i.e. ${\rm tr}[\hat A\hat B]=\sum_{ss'}[A_{ss'}B_{s's}]$. 
The bare one-photon current vertex is hence obtained
in terms of variational derivatives of non-interacting Green's function versus the gauge field:
\begin{eqnarray}
 \hat \Lambda^{(0)}_{1}(1,1';1'')
 &=&
  \frac{\delta   \hat G^{-1}_0(1,1';{\bf A})}{\delta A(1'')}\Big|_{\bf A\to 0}
  \nonumber\\
  &=&
  -ev\hat\sigma_y
  \delta(1-1')\delta(1-1'')
\label{bareL1}
  .
\end{eqnarray}

Because of the linear dispersion of the Dirac model,
the two- and three-photon current vertices are of course
null at the non-interacting level.
However, as we are going to see, the presence of a field-dependent self-energy $\hat\Sigma(1,1';{\bf A})$
in Eq. (\ref{eq:dyson}) breaks the linear dependence of the inverse
Green's function $\hat G^{-1}(1,1';{\bf A})$ on the external field,
and it is expected to give rise to higher order $n$-photon current vertices.
More precisely, such nonlinear processes can be computed
in terms of
multi-point correlation functions for $n$-photon vertex operators : 
\begin{align}
\hat{\Lambda}_n (1',1'';1,\dots,n) = \frac{1}{(n-1)!} \frac{\delta^n \hat G^{-1}(1',1'';{\bf A})}
{ \delta A(1),\dots,\delta A(n)  }\Big|_{\bf A\to 0}
.
\end{align} 
The possibility of obtaining a computationally affordable expression for  $\hat\Lambda_n (1',1'';1,\dots,n)$
depends of course on the specific characteristics of the scattering source.

In the following we focus on the role of elastic disorder/impurity scattering
which, along with a particularly simple structure allowing for a direct computation,
preserves in Dirac materials the fundamental frequency-dependence
of the self-energy, which is a crucial ingredient in determining the nonlinear electromagnetic response.
\subsection{One-particle impurity self-energy}
We consider scattering on local impurity centers with density $n_{\rm imp}$ and potential $V_{\rm imp}({\bf r})=\sum_iV_i\delta({\bf r}-{\bf R}_i)$
where ${\bf R}_i$ are the coordinates of the lattice sites.
We assume standard Born impurity correlations,
so that $\langle V_{\rm imp}({\bf r}) \rangle=0$ and the effective scattering potential reads 
\begin{align}\label{eq:Vpotential}
V(1,2) = \langle V_{\rm imp}({\bf r}_1) V_{\rm imp}({\bf r}_2)  \rangle_{\rm imp}   =
n_{\rm imp} V^2_{\rm imp}
\delta({\bf  r}_1-{\bf r}_2)~,
\end{align}
where the average $\langle \dots \rangle_{\rm imp}$ is meant over all the impurity configurations,
and where $V_{\rm imp}$ parametrizes the strength of impurity scattering.

In the absence of external fields,
the lowest-order self-consistent Born self-energy reads:
\begin{eqnarray}
\hat \Sigma(z) 
&=&
n_{\rm imp} V^2_{\rm imp}
\sum_{\bf k} \hat G({\bf k},z) 
\nonumber\\
&=&
\gamma_{\rm imp}
S_{\rm cell}  \int \frac{d^2k}{(2\pi)^2}\hat G({\bf k},z)~, 
\end{eqnarray}
where $\gamma_{\rm imp}=n_{\rm imp} V^2_{\rm imp}$,
$S_{\rm cell}$ is the two-dimensional unit-cell area,
and  the variable $z$ lies in the complex frequency space.
Due to the isotropic impurity scattering the self-energy spinor structure is trivial as  $\hat \Sigma(z)= \Sigma(z)\hat I$ and therefore the Green's function can be explicitly written as follows 
\begin{align}
\hat G({\bf k},z) = \frac{S(z) \hat I+ \hbar v 
\hat{\bm \sigma} \cdot{\bf k}}{S(z)^2-(\hbar v k)^2}~,
\end{align}
where where $S(z) = z+\mu_0-\Sigma(z)$. 

The introduction of a high-energy (ultra-violet) cut-off is an unavoidable requirement
of Dirac models. There is however a relative large degree of freedom in
the way how to introduce it, and particular care is needed in order to avoid
spurious results and to preserve physical consistencies, like Ward's identities, and gauge invariance. Dimensional regularization has proven to be a formidable tool
to ensure that physical correctness is preserved \cite{Leibbrandt_rmp_1975,Peskin}.
We consider the evaluation of the disorder self-energy which displays a primary diverging integral. In arbitrary D dimensions, we have thus
\begin{align}
\Sigma(z) = \frac{\gamma_{\rm imp}  N_{\rm cell} S^D_{\rm cell}}{(\hbar v)^D}  \int \frac{d^D \ell}{(2\pi)^D} \frac{S(z) }{S(z)^2-\ell^2}~.
\end{align}
Note that the above integral in $D$ dimensions can be solved in terms of Euler's Gamma-function, $\Gamma_E(z)$, by utilizing the following identity \cite{Peskin}
\begin{align}
\int \frac{d^D \ell}{(2\pi)^D} \frac{1}{(\ell^2+\Delta)^n} = 
\frac{1}{(4\pi)^{D/2}} \frac{\Gamma_E(n-\frac{D}{2})}{\Gamma_E(n)} 
\left(\frac{1}{\Delta}\right)^{n-D/2}~.
\end{align}
We set $D=d-\epsilon$ where $d=2$ is the physical dimension and $\epsilon\to 0$. 
Note that $\Gamma_E(\epsilon/2) \approx  2/\epsilon$ for  $\epsilon\to 0$ and
\begin{align}
\lim_{\epsilon \to 0} \frac{(X^2)^{-\epsilon/2}}{\epsilon/2} = \ln \left[\frac{W^2}{X^2}\right]~,
\end{align}
where $W$ is a proper ultra-violet energy cut-off,
and where
we use the prescription $\lim_{\epsilon \to 0} 1/\epsilon \equiv \ln[W]$.

Eventually, we obtain the following self-consistent formula for the self-energy 
\begin{align}\label{eq:self}
\Sigma (z) = - U S(z)\ln\left[- \frac{W^2}{S(z)^2} \right]~, 
\end{align}
where $U$ is a dimensionless parameter representing the electron-impurity scattering strength: 
\begin{align}
 U = \frac{ \gamma_{\rm imp} S_{\rm cell}}{4\pi (\hbar v)^2}~. 
\end{align}
In a direct comparison with graphene,
we have $S_{\rm cell} = \sqrt{3}a^2/2$ and $\hbar v = 3 a t_0/2$,
where $a\approx 0.246$ nm is the lattice constant and $t_0\sim3$eV is the nearest neighbor hopping energy.  
In order to preserve the number of states, $S_{\rm cell}$ defines also
an effective 2D Brillouin zone, $V_{\rm BZ}=4\pi^2/S_{\rm cell}$ which, employing
the isotropic symmetry, defines a momentum cut-off $k_c$, $\pi k_c^2=V_{\rm BZ}$,
and a natural ultra-violet energy cut-off $W=\hbar v k_c$ for the Dirac linear dispersion. 
Using the above parameters for graphene we get $W=7.2$ eV.

The above procedure of dimensional regularization is employed in similar way
in the evaluation of all the momentum integrals in the present work. 
\section{Many-body-driven multi-photon vertex generation} \label{sec:KB}
The analytical expression of the functional dependence of the one-particle self-energy
on the external field allows, within the spirit of a Baym-Kadanoff approach,
the derivation of a closed set of self-consistent equations governing the transport properties
at the chosen (linear or nonlinear) order.
For Dirac materials, like graphene, where the second order response vanishes
by symmetry, a particular interest is paid
on the {\em third-order} response, and, within this framework, on the third-harmonic generation.
At the non-interacting level, vertex corrections  are null and the third-harmonic generation is governed
by the well-known ``square'' diagram with bare one-photon current vertices at the corners \cite{Rostami_prb_2016}.

Things are much more complex in the presence of many-body interactions where
the intrinsic dependence of the self-energy on the frequency and on external fields
triggers in novel nonlinear effects which are not predictable
at the non-interacting level or within a phenomenological model using a constant
(frequency-independent and external-field-independent) one-particle scattering rate.

A careful investigation of the many-body effects driven by disorder scattering,
at the lowest-order self-consistent Born level,
is remarkably enlightening since it preserves all the relevant nonlinearity
but with a particular simple expression for the self-energy
which results to depend linearly on the fully interacting Green's function
in the presence of external field:
\begin{align}\label{eq:scba}
  \hat \Sigma(1,2) = V(1,2)   \hat G(1,2)~,
\end{align}
Here, $V(1,2)$ stands for the many-body interaction potential where for the impurity-driven the interaction potential is given by Eq. (\ref{eq:Vpotential}).

\begin{figure*}[t]
\begin{overpic}[width=0.9\textwidth]{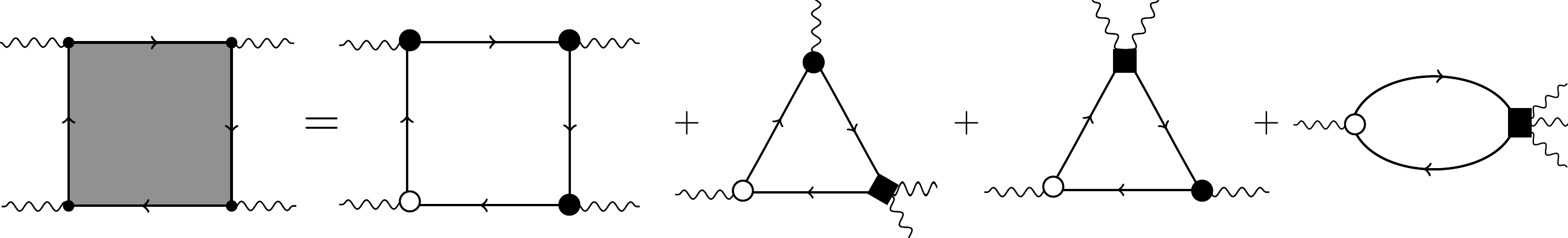} 
\end{overpic}
\caption{Diagrams for the third-order response function of graphene as a two-dimensional Dirac materials in terms of {\em renormalized} one-, two- and three-photon vertices (where $n$ is given by the number of the attached wavy lines (photons). The empty circle
represents the one-photon current vertex at the non-interacting level [Eq. (\ref{bareL1})],
whereas filled symbols represent fully renormalized $n$-photon vertices.}
\label{fig:chi3}
\end{figure*}

On this basis,
after performing a lengthy but straightforward algebra, we can construct a diagrammatic theory
for the third-order response function of graphene as a two-dimensional Dirac material.
Feynman diagrams for the third harmonic response function
$\rchi^{(3)}(1;2,3,4)$
are depicted in Fig.~\ref{fig:chi3}. 
Here solid lines represent Green's functions, wavy lines the incoming/outcoming photons,
and the empty/filled symbols the bare/renormalized $n$-photon vertices, where $n$ can be identified
by the number of attached wavy-lines (photons).

A key role in this context is played by the multi-photon ($n > 1$) current vertices $\hat\Lambda_n (1',1'';1,\dots,n)$.
A close inspection reveals that each fully dressed $n$-photon vertex $\hat\Lambda_n (1',1'';1,\dots,n)$ can be expressed
in the self-consistent (Bethe-Salpeter-like) form (Fig. \ref{fig:vertex_BS}):
\begin{figure}[t]
\begin{overpic}[width=0.4\textwidth]{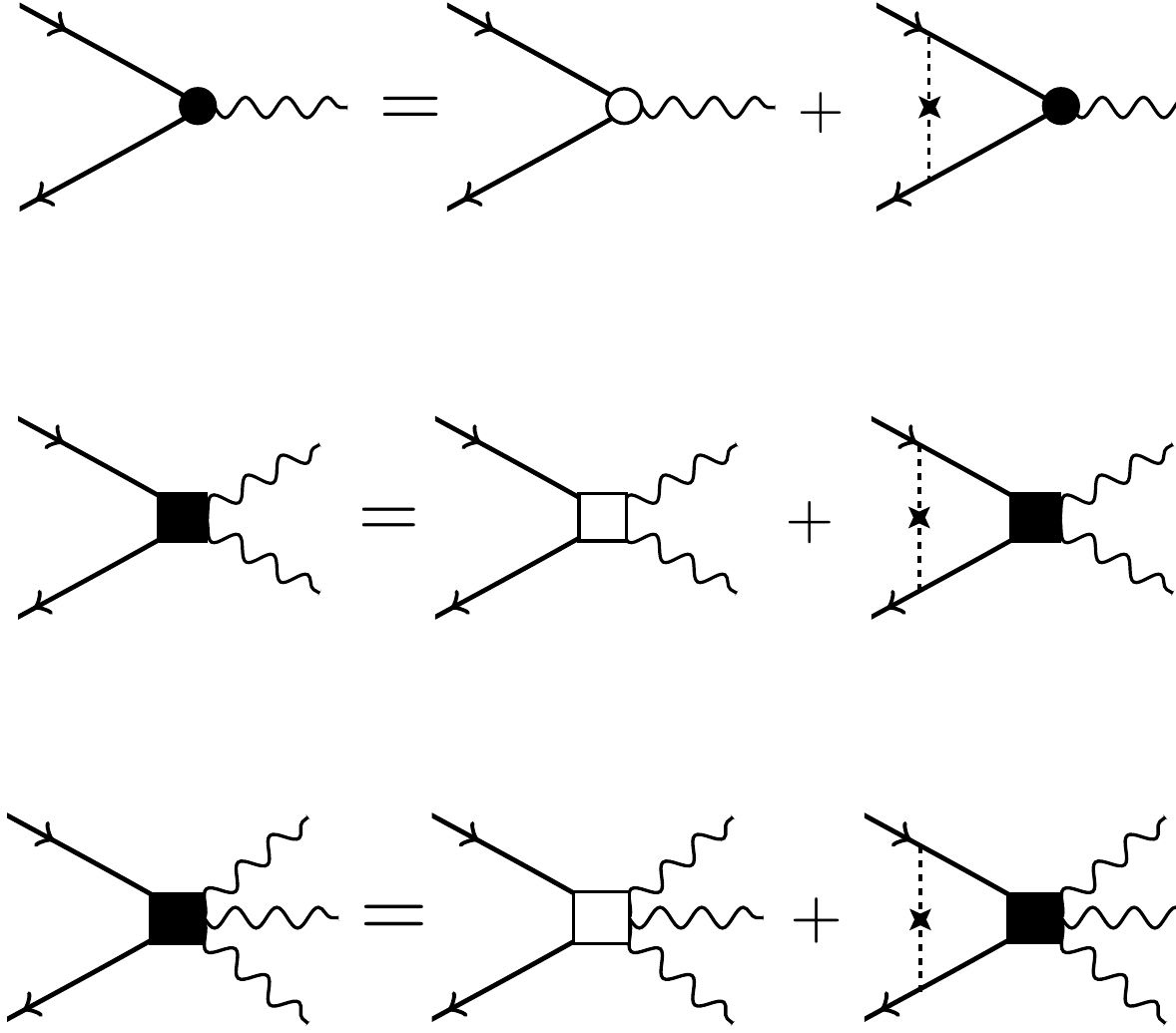} 
\put(-7,77){(a)}
\put(-7,42){(b)}
\put(-7,8){(c)}
\end{overpic}
\caption{Diagrams for the self-consistent Bethe-Salpeter renormalization of $n$-photon vertex functions, $\Lambda_n$, where panels (a), (b) and (c) corresponds to $n=1$, $n=2$ and $n=3$, respectively.}
\label{fig:vertex_BS}
\end{figure} 
\begin{eqnarray}
&&\hat\Lambda_n (1',1'';1,\dots,n)
=
\hat\lambda_n (1',1'';1,\dots,n)
\nonumber\\
&&
+
\int_{\bar 2,\bar 3}
\hat{K}_n(1',1'';\bar 2,\bar 3)
\hat\Lambda_n (\bar 2,\bar 3;1,\dots,n),
\end{eqnarray}
where the term $\hat\lambda_n (1',1'';1,\dots,n)$ (empty symbols in Fig.~\ref{fig:vertex_BS})
can be expressed in terms of {\em lower} order multi-photon vertices
(see Fig.  \ref{fig:vertex_0}).
\begin{figure}[t]
\begin{overpic}[width=0.3\textwidth]{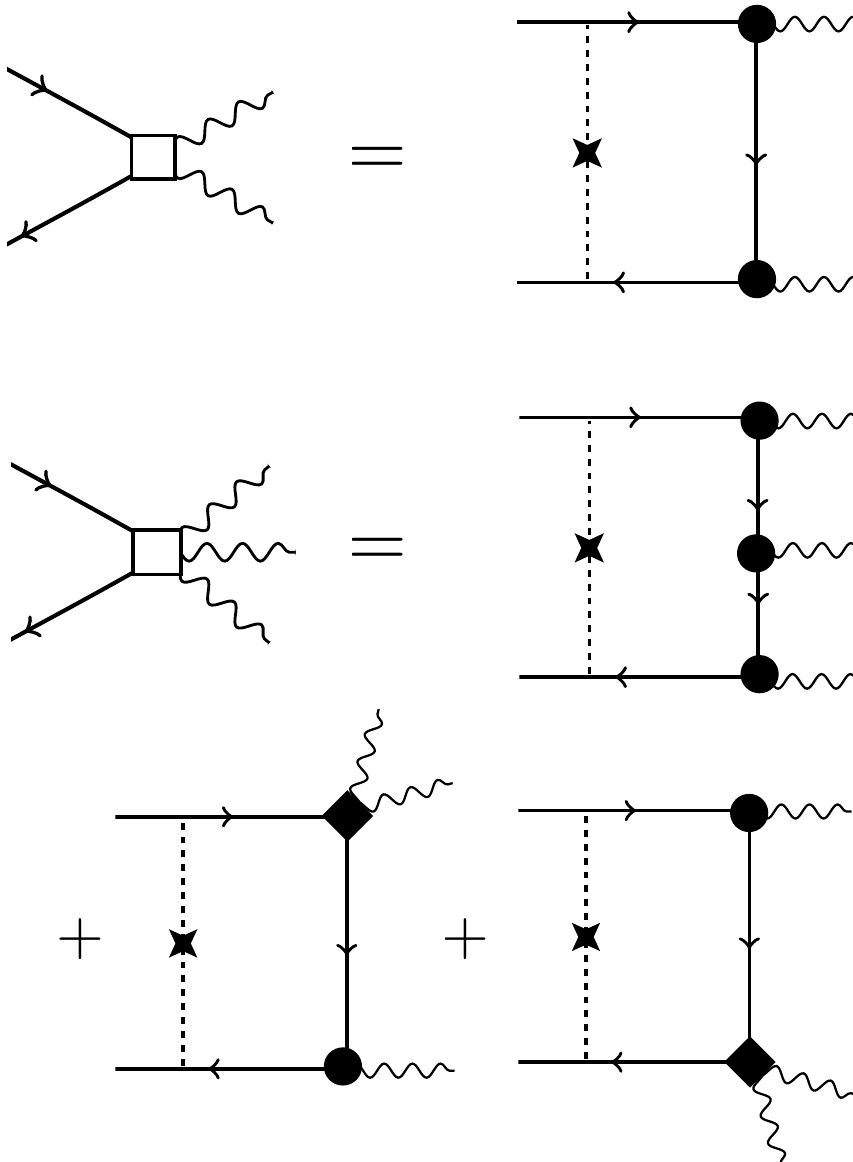} 
\put(-10,85){(a)}
\put(-10,50){(b)}
\end{overpic}
\caption{Diagrams for the interaction induced two- and three-photon vertices, $\lambda_{n=2,3}$ , are depicted in panels (a) and (b), respectively.}
\label{fig:vertex_0}
\end{figure} 
It should be noticed that, while $\hat\lambda_1 (1',1'';1)$ reduces to 
the bare one-photon vertex $\hat\lambda_1 (1',1'';1)=\hat\lambda_1^{(0)} (1',1'';1)
= -ev\hat\sigma_y  \delta(1-1')\delta(1-1'')$
 in the non-interacting limit $U \rightarrow 0$,
the two- and three-photon vertices terms, $\hat\lambda_2 (1',1'';1,\dots,n)$, $\hat\lambda_3 (1',1'';1,\dots,n)$
and so the fully dressed multi-photon vertices $\hat\Lambda_2 (1',1'';1,\dots,n)$, $\hat\Lambda_3 (1',1'';1,\dots,n)$
are triggered in by the many-body impurity scattering.
By using the symmetry enforced by isotropic impurity scattering, one can see
that each $n$-photon vertex has a specificed Pauli structure.
This property can be employ to define a scalar vertex function
for each $n$-photon vertex, namely:
$\hat \Lambda_n= (-ev\hat\sigma_y)^n \Lambda_n$,
and
$\hat \lambda_n= (-ev\hat\sigma_y)^n \lambda_n$,
With this notation in the non interacting case we have
$\Lambda_1^{(0)}=\lambda_1^{(0)}=1$ and $\Lambda^{(0)}_{n>1}=\lambda^{(0)}_{n>1}=0$.
\section{Many-body dressed third harmonic generation} \label{sec:Pfunction}
The diagrammatic expressions  in Figs. \ref{fig:chi3}-\ref{fig:vertex_0}
are valid for any generic third-order optical response.
A particularly interesting case is the third-harmonic generation (THG),
which, using the translational invariance symmetry
and in the Matsubara space,
can be conveniently written as:
\begin{align}
\rchi^{(3)}_{\rm THG} (m) =   \frac{1}{\beta}\sum_{n} P(n,n+m,n+2m,n+3m).
\end{align}
Here $m=i\omega_m$ represents the photon bosonic energy,
and $n=i\omega_n$ is the internal fermionic energy to be summed over.

The corresponding third-harmonic optical conductivity $\sigma^{(3)}_{\rm THG}(\omega)$
can be hence obtained after analytical continuation $i\omega_m\to \hbar\omega+i0^+$ as:
\begin{align} \label{eq:sigma3}
\sigma^{(3)}_{\rm THG}(\omega)
=  i \frac{\rchi^{(3)}_{\rm THG}(\omega) }{\omega^3}~.
\end{align}

After a straightforward algebra one can obtain thus \cite{Rostami_Cappeluti_arxiv_2020}
\begin{align}\label{eq:chi3}
\rchi^{(3)}_{\rm THG}(\omega) &= 
 \int^{+\infty}_{-\infty} \frac{d\epsilon}{2\pi i} 
\bigg\{ 
n_{\rm F}(\epsilon) P^{\rm RRRR}
-n_{\rm F}(\epsilon+3\hbar\omega) P^{\rm AAAA} 
\nonumber\\&+  [n_{\rm F}(\epsilon+\hbar\omega) 
-n_{\rm F}(\epsilon)  ]  P^{\rm ARRR}
\nonumber\\&
+  [n_{\rm F}(\epsilon+2\hbar\omega) 
- n_{\rm F}(\epsilon+\hbar\omega) ] P^{\rm AARR}
\nonumber\\&+ [n_{\rm F}(\epsilon+3\hbar\omega) 
-n_{\rm F}(\epsilon+2\hbar\omega) ] P^{\rm AAAR}
\bigg\}.
\end{align}
For sake of compactness, we use here a short notation
where
$P^{\nu_0\nu_1\nu_2\nu_3}=
P(\epsilon_{0,\nu_0},\epsilon_{1,\nu_1},\epsilon_{2,\nu_2},\epsilon_{3,\nu_3})$,
where
$\epsilon_{j,\nu_j}=\epsilon+j \omega+j\eta_{\nu_j}$ ($j=0,1,2,3$)
and where $\eta_{\nu_j}$ is a vanishingly small quantity with
$\eta_{\nu_j}>0$ if $\nu_j={\rm R}$ and $\eta_{\nu_j}<0$ if $\nu_j={\rm A}$.
Notice also that $P^{\rm AAAA} = (P^{\rm RRRR})^\ast$.

As depicted in Fig.~\ref{fig:chi3}, the $P$-function contains four  
different contributions, $P= P_1+P_2+P_3+P_4$,
associated respectively with square ($P_1$), triangles ($P_2$, $P_3$) and bubble ($P_4$) diagrams.
More explicitly we can write:
 \begin{align}
P_{1}(z_0,z_1,z_2,z_3) &
= \alpha
 Q_1(z_0,z_1) 
 Q_1(z_1,z_2)
  Q_1(z_2,z_3)
  \nonumber\\&\times
 \Omega_1(z_0,z_1,z_2,z_3)~.
 \label{funcP1}
 \end{align}
where $\alpha={e^4 v^2 N_f}/{2\pi \hbar^2}$. 
The sum over spin and valley indices just leads to an overall degeneracy factor $N_f=N_s N_v$ where $N_s=2$ and $N_v=2$.
The function $\Omega_1(z_0,z_1,z_2,z_3)$ represents the square diagram neglecting
vertex renormalization,
 \begin{eqnarray}
 \Omega_1(z_0,z_1,z_2,z_3) &=\frac{\displaystyle \gamma_{\rm imp}}{\displaystyle 2U}
\sum_{{\bf k}} {\rm Tr} [\hat\sigma_y \hat G({\bf k},z_0)\hat\sigma_y  \hat G({\bf k},z_1)\hat\sigma_y  \nonumber\\&\times 
 \hat G({\bf k},z_2)  \hat \sigma_y \hat G({\bf k},z_3)]
 ,
 \end{eqnarray}
and $Q_1(z_i,z_j)=\Lambda_1(z_i,z_j)/ \lambda_1(z_i,z_j)$,
where as defined above $\lambda_1(z_i,z_j)=1$.
$Q_1(z_i,z_j)$ represents thus the one-photon Bethe-Salpeter renormalization factor 
which is discussed in details in Appendix \ref{app:vertex_1}.
In similar way we can write the contributions 
 of the two triangles diagrams as
 \begin{align}
P_{2}(z_0,z_1,z_2,z_3) & = 
\alpha
Q_1(z_2,z_3)Q_2(z_0,z_2)
\lambda_2(z_0,z_1,z_2) \nonumber\\&\times \Omega_2(z_0,z_2,z_3)
,
\end{align}
where  $\lambda_2(z_0,z_1,z_2)$ is the lowest order two-photon current vertex (Fig. \ref{fig:vertex_0}a),
 \begin{equation}
\Omega_2(z_0,z_2,z_3) 
=
-\frac{\gamma_{\rm imp}}{2U}
\sum_{{\bf k}}{\rm Tr} [\hat\sigma_y \hat G({\bf k},z_0) \hat G({\bf k},z_2)\hat \sigma_y \hat G({\bf k},z_3)]~,
 \end{equation}
and $Q_2(z_i,z_j)=\Lambda_2(z_i,z_k,z_j)/\lambda_2(z_i,z_k,z_j)$
is two-photon Bethe-Salpeter renormalization factor (see Appendix \ref{app:vertex_2}). Furthermore, we can also express the triangle diagram as:
 \begin{eqnarray}
P_{3}(z_0,z_1,z_2,z_3)
&=& 
\alpha
Q_1(z_0,z_1)Q_2(z_1,z_3)
\lambda_2(z_1,z_2,z_3)  \nonumber\\
&&\times  \Omega_3(z_0,z_1,z_3)~,
\end{eqnarray}
where
 \begin{align}
 \Omega_3(z_0,z_1,z_3)
 =
  - \frac{ \gamma_{\rm imp}}{2U}
\sum_{{\bf k}} {\rm Tr} [\hat\sigma_y \hat G({\bf k},z_0)  \hat\sigma_y \hat G({\bf k},z_1) \hat G({\bf k},z_3)]~.
 \end{align}

Finally we can express the bubble term as: 
\begin{eqnarray}
P_{4}(z_0,z_1,z_2,z_3)
&=&
\alpha  \lambda_3(z_0,z_1,z_2, z_3) 
Q_3(z_0,z_3)
\nonumber\\
&&
\times
X_1(z_0,z_3)
,
\end{eqnarray}
where $Q_3(z_i,z_k,z_l,z_j)=\Lambda_3(z_i,z_j)/\lambda_3(z_i,z_k,z_l,z_j)$
is three-photon Bethe-Salpeter renormalization factor (see Appendix \ref{app:vertex_3}),
with $\lambda_3(z_0,z_1,z_2, z_3)$ being the lowest order three-photon vertex function (Fig. \ref{fig:vertex_0}b),
and
\begin{align}
 X_1(z,z') = \frac{\gamma_{\rm imp}}{2 U} \sum_{\bf k} {\rm Tr}[\hat\sigma_y \hat G({\bf k},z)\hat\sigma_y \hat G({\bf k},z')]~.
\end{align}

 A close inspection of the topological structure of diagrams for the three-photon vertex (see Fig.~\ref{fig:vertex_0}b and Appendix \ref{app:vertex_3})
permits further simplifications as:
\begin{eqnarray}
P_{4}(z_0,z_1,z_2,z_3) 
&= &
U Q_3(z_0,z_3) X_1(z_0,z_3) 
\nonumber\\
&&
\times
\sum^3_{i=1}P_{i}(z_0,z_1,z_2,z_3),
\end{eqnarray}
Using $Q_3 =Q_1=1/[1-UX_1]$ (see Appendix \ref{app:vertex_3}),  we find the following result for the total $P$-function 
\begin{align}
P(z_0,z_1,z_2,z_3) = Q_3(z_0,z_3)  \sum^3_{i=1}P_{i}(z_0,z_1,z_2,z_3). 
\end{align}
We can see thus that the net impact of three-photon vertex diagram in the third-order response function
simply leads to the appearance of
the three-photon renormalization factor $Q_3(z_0,z_3)$
on the contribution of  the other diagrams.
The explicit expressions of  $\lambda_n$, and $Q_n$, and $ \Omega_n$
are provided in great details in Appendix
sections for the one-, two-, and three-photon vertex Bethe-Salpeter renormalizations. 

Equipped with the all analytical expressions needed for the computation of optical properties of third-harmonic generation
response, in the following Section we
present numerical results for the low-energy intraband third-harmonic conductivity of
two-dimensional Dirac modelling of graphene.
 \begin{figure}[t]
\begin{overpic}[width=0.43\textwidth]{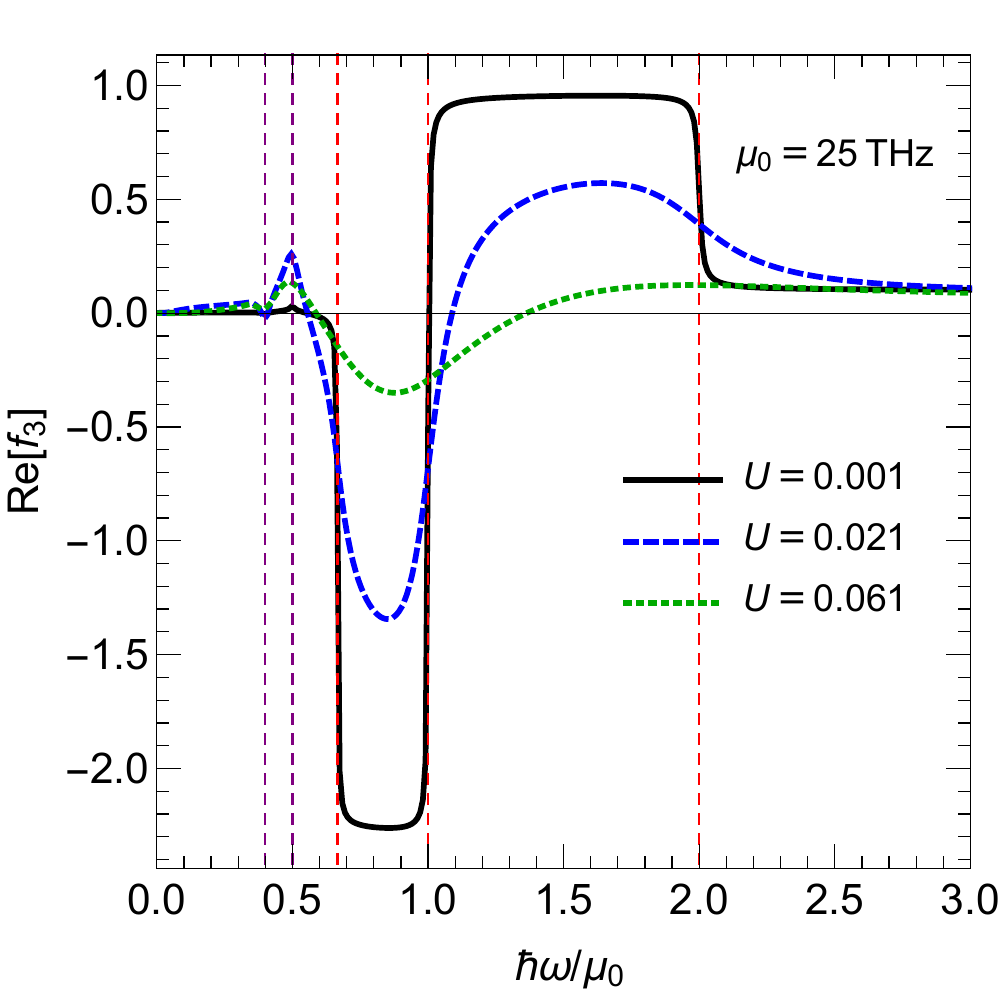} 
\put(77,20){\large(a)}
\put(25,95){\rotatebox{90}{ n=5}}
\put(28,95){\rotatebox{90}{ n=4}}
\put(32.5,95){\rotatebox{90}{ n=3}}
\put(41.5,95){\rotatebox{90}{ n=2}}
\put(68.5,95){\rotatebox{90}{ n=1}}
\end{overpic}
\begin{overpic}[width=0.43\textwidth]{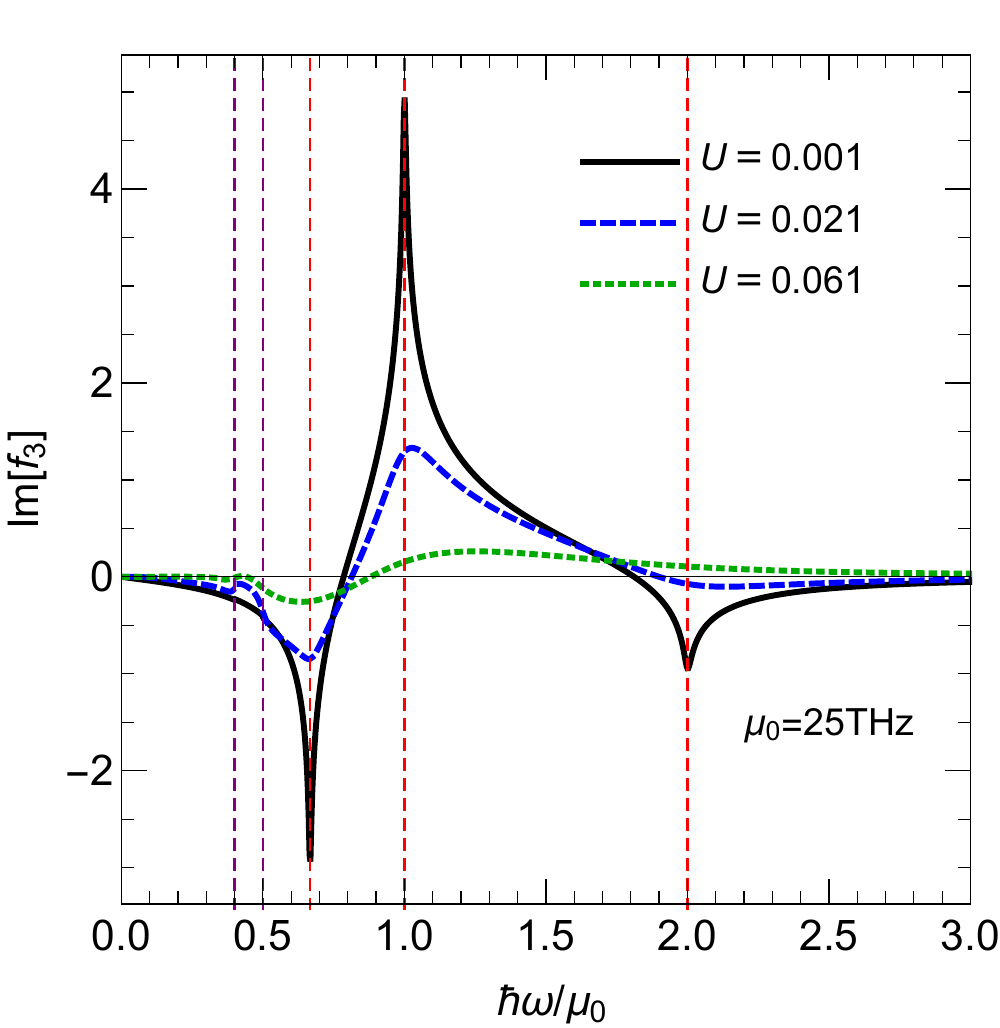} 
\put(50,20){\large(b)}
\end{overpic}
\caption{Real (panel a) and imaginary (panel b) parts of the THG conductivity for the whole range of intra to inter-band regimes.
The vertical red dashed lines indicate the location of $n$-photon inter-band resonance as $\hbar\omega = 2\mu_0/n$ with $n=1,2,3$,
whereas the vertical purple dashed lines reveals the frequencies of four and five-photon incoherent-transitions with $n=4,5$.}
\label{fig4}
\end{figure}
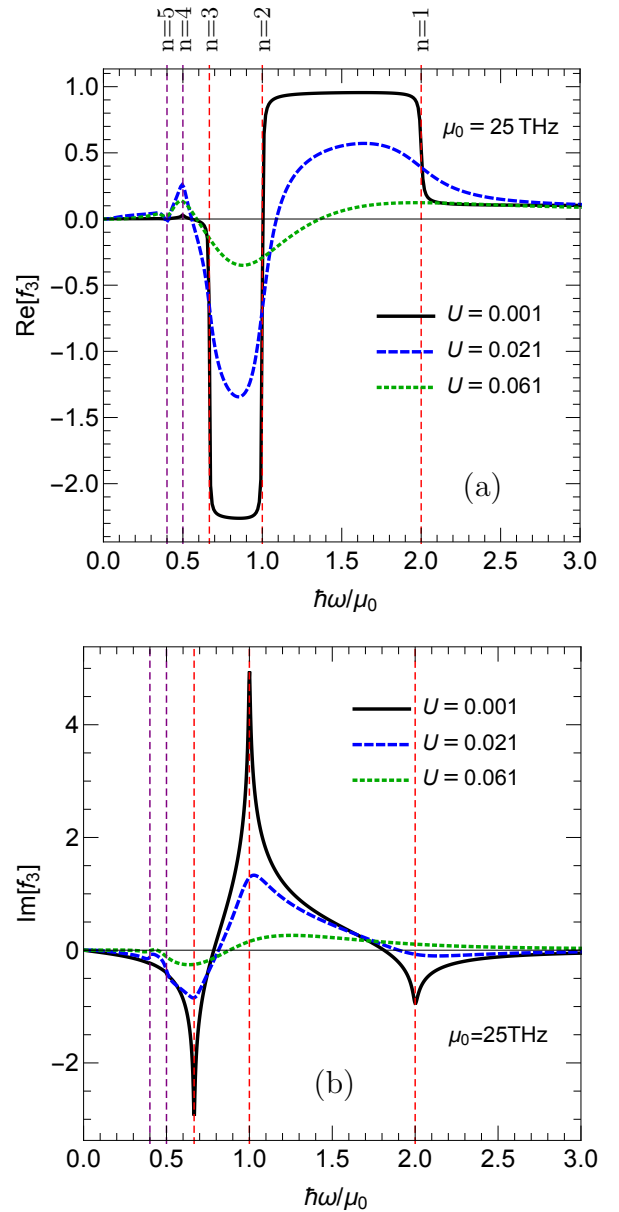 

\section{Results and Discussion}\label{sec:result_discussion}
Based on a mere dimensional analysis, we can conveniently express
the  zero-temperature third harmonic conductivity of graphene in terms of a dimensionless
function $f_3(x,y,z)$:   
\begin{align}
 \sigma^{(3)}_{\rm THG} (\omega) = \frac{\sigma_0}{E^2_0} \left(\frac{t_0}{\hbar\omega} \right)^4 f_3\left(\frac{\hbar\omega}{\mu_0},
\frac{\mu}{\Gamma},U\right)
\end{align}
where 
$\mu=\mu_0-\mbox{Re}\Sigma(0)$ is the renormalized chemical potential,
 $\Gamma = \Gamma(0)=-\mbox{Im}\Sigma(0)$  is the Fermi surface scattering rate,
$\sigma_0 = e^2/4\hbar$ is the universal conductivity including the spin and valley degeneracy,
and $ E_0 =  \pi t_0/\sqrt{3} ea\approx 22.0~{\rm V/nm}$ is a characteristic electric-field scale
determined by the inter-atomic hopping energy $t_0$ and by the lattice constant $a$.
In the dc limit $\omega \to 0$, $\lim_{x\to0}f_3(x,y,z)/x^4$, we recover
the transport regime discussed in Ref.~\cite{Rostami_Cappeluti_arxiv_2020}.  
The appearance of the bare chemical potential $\mu_0$
in the definition of $x$, and of the renormalized one $\mu$ in $y$
is dictated by the different role of $\mu_0$ and $\mu$ in governing the dc and optical properties,
as discussed in more details below.

Besides the obvious role of the real part of the third harmonic generation,
the imaginary part of the conductivity bares also a strong relevance. 
As a matter of fact, the THG efficiency $\eta_{\rm THG}$ scales indeed
as $\eta_{\rm THG} = I_{\rm THG}/I_{\rm in} \propto I^2_{\rm in} |\sigma^{(3)}_{\rm THG}|^2$ where $I_{\rm in}$ and $I_{\rm THG}$ stand for the incident and THG intensities, respectively. 
For computational reasons it is convenient to calculate by using of  Eqs.~(\ref{eq:sigma3})-(\ref{eq:chi3})
 directly the real part of the nonlinear conductivity ${\rm Re}[\sigma^{(3)}_{\rm THG}]$,
or equivalenty ${\rm Re}[f_3] \propto \omega {\rm Im}[\rchi^{(3)}_{\rm THG}]$.
The imaginary part of ${\rm Im}[\sigma^{(3)}_{\rm THG}]$ (or ${\rm Im}[f_3] \sim \omega {\rm Re}[\rchi^{(3)}_{\rm THG}]$) can be
thus obtained by means of the Kramers-Kronig relations:
\begin{align}
{\rm Re} [\rchi^{(3)}_{\rm THG} (\omega) ] =  \frac{2}{\pi} 
\int^\infty_0  d\omega'
\frac{ \omega' {\rm Im} [\rchi^{(3)}_{\rm THG} (\omega') ]}{\omega^2- \omega'^2}~.
\end{align}
 %

At high-frequency regime, the third-harmonic generation response function $\rchi^{(3)}_{\rm THG} (\omega)$ is described by {\em purely interband} transitions, $\rchi^{(3)}_{\rm THG,inter}(\omega)$, giving rise
to step-like functions at $\hbar\omega \approx 2 \mu_0/n$
corresponding to $n=1,2,3$ photon resonances where the three-photon resonance at $\hbar\omega \approx 2 \mu_0/3$ defines the interband optical edge. The low frequency $\hbar\omega\ll\mu$ corresponds to the {\em purely intraband} regime of third-harmonic optical conductivity. In addition to this structure, in the intermediate frequency range, here, we obtain novel incoherent four and five-photon transitions with {\em mixed intra-interband} characters. 

\subsection{Four and five-photon incoherent transitions } 

The full quantum treatment of many-body interaction with Green's functions and Kubo formalism
predicts intriguing novel spectral features.
In Fig.~\ref{fig4}a,b we plot the real and imaginary parts of the THG function $f_3 \sim \omega^4  \sigma^{(3)}_{\rm THG}$ versus the laser-frequency $\omega$ in the whole intra- to inter-band frequency ranges  for different values of the scattering strength $U$.  In the almost clean limit ($U=0.001$) the dominant features are the
the multi-photon interband resonances at $\hbar\omega \approx 2 \mu_0/n$ with $n=1,2,3$, which appear
as smeared structures by the interaction in the real and imaginary parts.
Note that on this scale the purely intraband features are not visible  since their magnitude
scales as $\omega^4$. Upon increasing the scattering strength, besides the obvious smearing of the interband features, we notice the appearance of novel resonances below the interband edge $\hbar\omega \approx 2 \mu_0/3$,
i.e. in the intraband range. A closer look reveals that such spectral features occurs at frequencies $\hbar\omega = 2\mu_0/n$ with $n=4,5$ (purple vertical dashed lines in Fig. \ref{fig4}), namely at energies corresponding to four- and five-photon resonances, respectively. 

The very possibility of observing four- and five-photon transitions in the third order conductivity
is quite surprising and it calls for a further deeper investigation.
Also worth being noticed is the fact that, although a full many-body treatment is here enforced,
the $n$-photon resonances ($n=1,\ldots,5$) occurs at the energies $\hbar\omega \approx 2 \mu_0/n$
dictated by the {\em bare} chemical potential $\mu_0$, rather than by
the effective renormalized one $\mu$.
This puzzling result has not been detected previously in literature
since in non-interacting models as well as in phenomenological models where
only a constant scattering rate $\Gamma$ (imaginary part of self-energy) is included,
no renormalization of the chemical potential is operative, and $\mu=\mu_0$.

\begin{figure*}[t]
\hspace{4mm}
\begin{overpic}[width=0.385\textwidth]{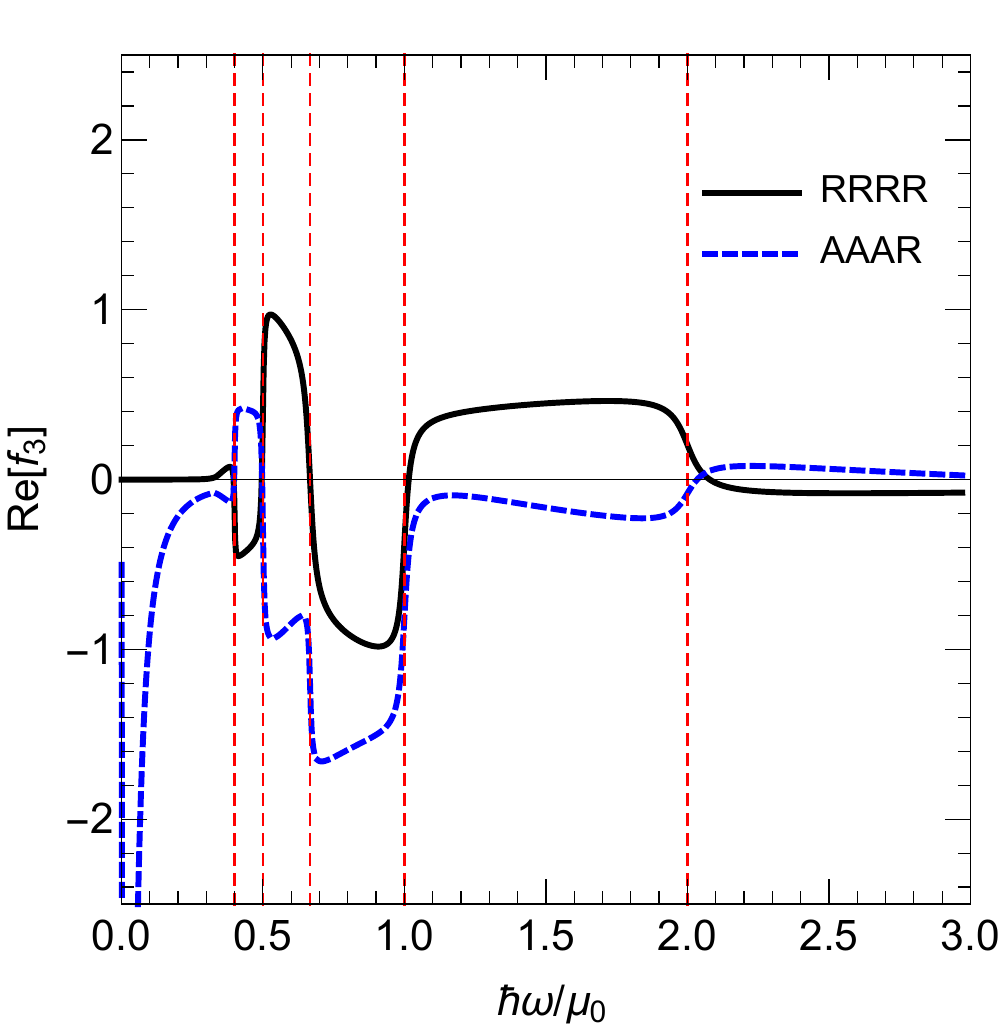} 
\put(14,87){\large(a)}
\end{overpic}
\hspace{2mm}
\begin{overpic}[width=0.405\textwidth]{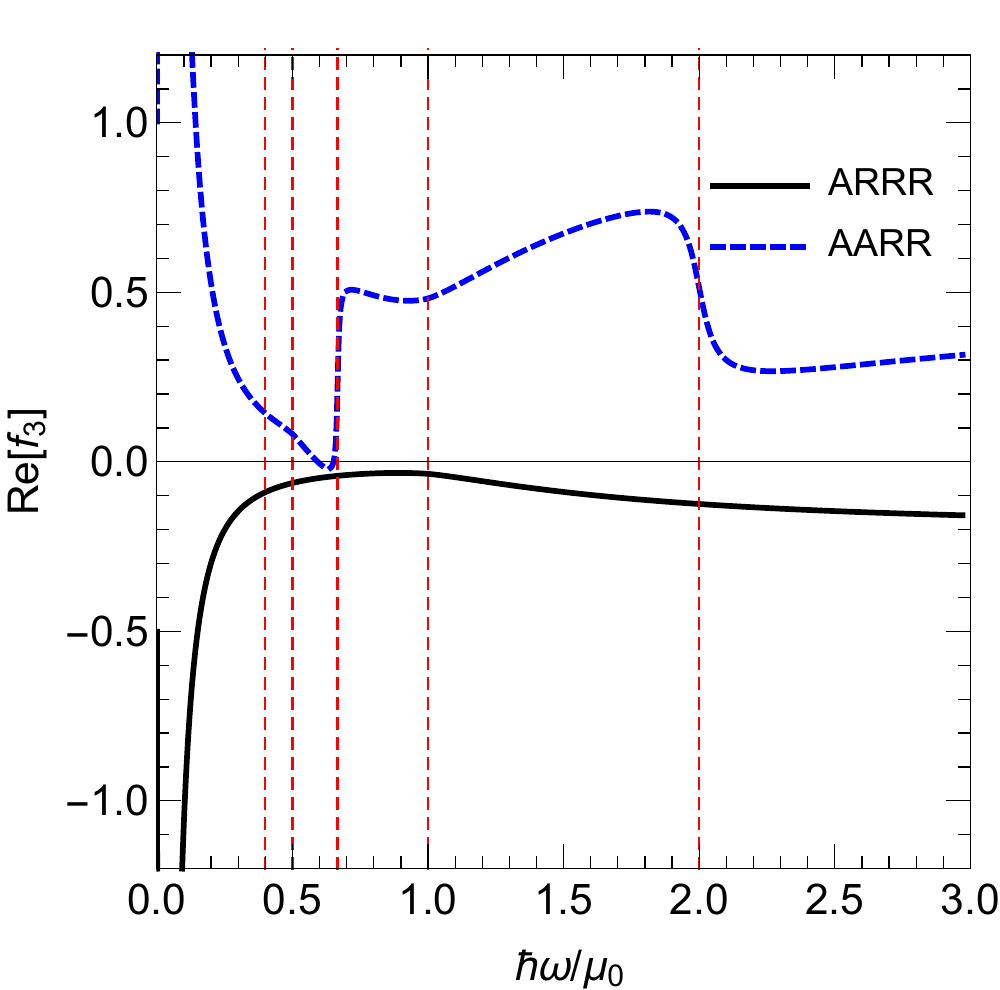} 
\put(85,85){\large(b)}
\end{overpic}
\begin{overpic}[width=0.395\textwidth]{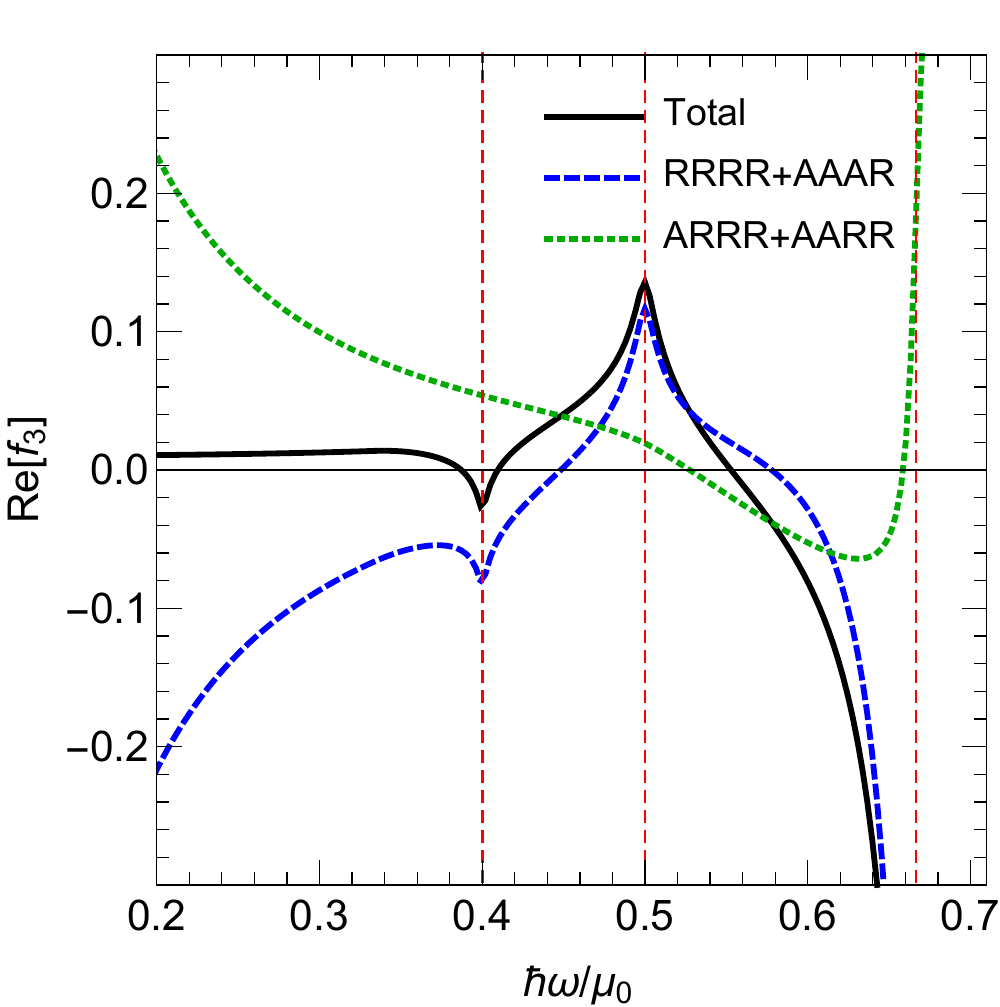} 
\put(19,86){\large(c)}
\end{overpic}
\hspace{3mm}
\begin{overpic}[width=0.39\textwidth]{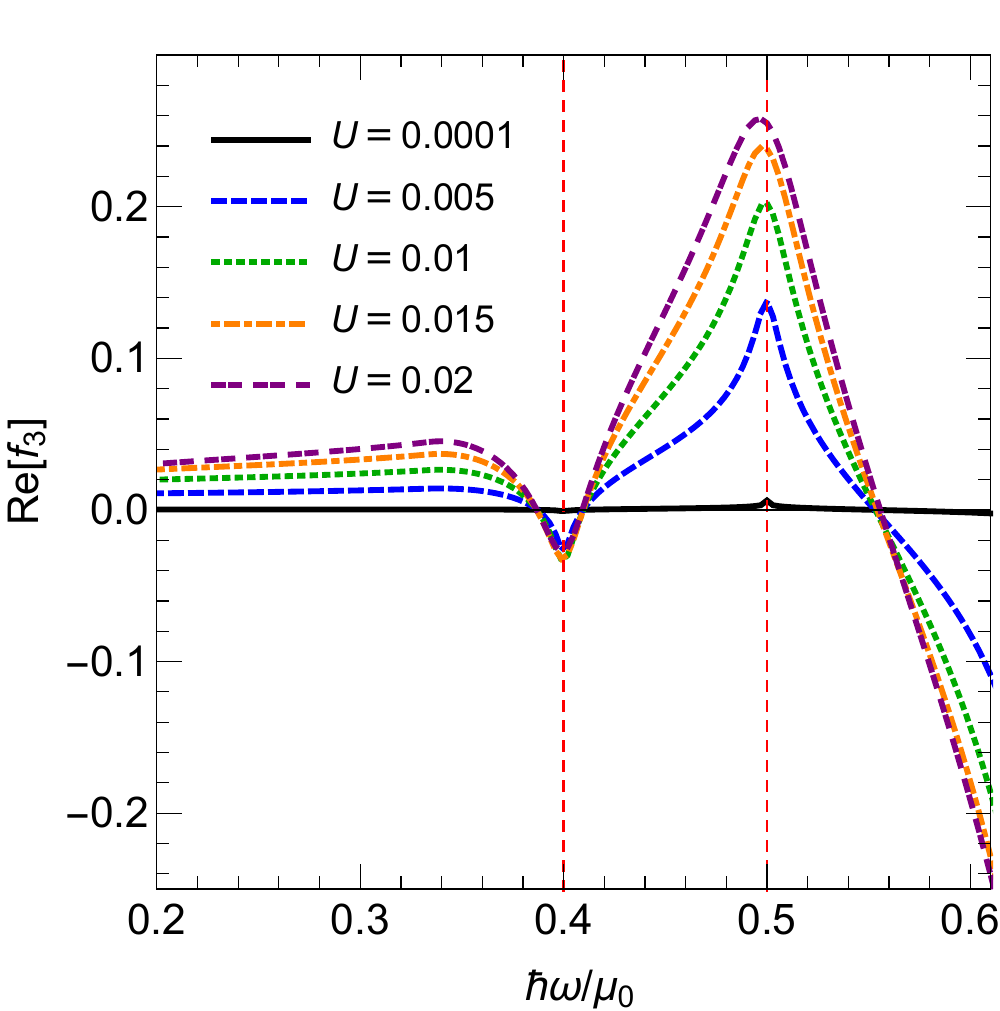} 
\put(86,86){\large(d)}
\end{overpic}
\caption{Panels (a)-(c): different channel contributions (RRRR, AAAR, ARRR, AARR) to the real part of $f_3$ function
for $U=0.005$. Panel (d): evolution of the real part of total $f_3$ function for different values of $U$.  
We set $\mu_0=25$THz.}
\label{fig:PXXXX}
\end{figure*} 

The appearance of four- and five-photon transitions in the third order conductivity can be rationalized
by considering at the simplest level the square diagrams depicted in Fig. \ref{fig:chi3}
furthermore neglecting the Bethe-Salpeter vertex  renormalization.
In this case, according Eq. (\ref{funcP1}), the kernel response function $P$ in Eq, (\ref{eq:chi3})
will read simple
$P(z_0,z_1,z_2,z_3)=P_{1}(z_0,z_1,z_2,z_3) \propto \Omega_1(z_0,z_1,z_2,z_3)$.
The explicit expression of $\Omega_1(z_0,z_1,z_2,z_3)$ is long and cumbersome
and it is provided in Appendix \ref{app:vertex_3}. The main feature in regards with the present issue
is that it depends as:
\begin{eqnarray}
 \Omega_1(z_0,z_1,z_2,z_3)
 &\propto&
 \Pi_{i\neq j=0,\ldots,3}
 \frac{1}{S(z_i)+S(z_j)}
.
\label{oureq}
\end{eqnarray}
Accordingly Eq. (\ref{oureq}), 
multi-photon transitions can occur when
${\rm Re}[S(z_i)+S(z_j)]=0$ with $z_j = \epsilon+j \hbar\omega$.  Neglecting for the moment the contribution of the self-energy,
this implies that $\epsilon=-\mu_0 - (i+j)\hbar\omega/2$.
We assume for simplicity $\mu>0$, $T=0$ and we focus on the RRRR channel.
Enforcing the boundary conditions $-3\hbar\omega \le \epsilon \le 0$  which originates from the  factor ``$n_{\rm F}(\epsilon)-n_{\rm F}(\epsilon+3\omega)$'' in Eq.~(\ref{eq:chi3}),
we get that possible $n$-photon inter-band transitions can occur at $ \hbar\omega = 2\mu_0/n $ with $n=6-(i+j)$,
and, considering $i,j=0,1,2,3$ with $i\neq j$, we obtain the possible values $n=1,2,3,4,5$.
Similar analysis can be applied for other channels 
where the possibility of detecting $n$-photon transition is dictated by the integration window over $\epsilon$  (see Fermi function prefactors in Eq.~(\ref{eq:chi3}) for different channels)
and by the retarded/advanced character of the complex frequencies $z_i$ involved
is the transition. More explicitly, it can be shown than 
$n$-photon transitions in each channel have a finite spectral weight only when $z_i$, $z_j$ have the {\em same} (retarded or advanced)
character, in similar way as it occurs in the linear optical response.
With such roadmap, we can analyze theoretically the possible
appearance of multi-photon transitions in each separate contribution
$P^{\rm RRRR}$, $P^{\rm ARRR}$, $P^{\rm AARR}$, and $P^{\rm AAAR}$.
Our theoretical predictions are summarized  in Table.~\ref{tab:n_photon},
and the numerical results are shown in Fig.~\ref{fig:PXXXX}a,b,
in excellent agreement with each other.

\begin{table}[ht]
\caption{Predicted $n$-photon transitions $ \hbar\omega = 2\mu_0/n $  in the third-order optical conductivity.
The transitions marked with asterix are expected to have null spectral weight the for elastic scattering here considered
but they might gain a finite spectral weight in the presence of inelastic scattering (see discussion in the text).}
\begin{center}\label{tab:n_photon}
\scalebox{1.1}{
 \begin{tabular}{c|c|c } 
 \hline \hline
Channel & $\epsilon$-range at $T=0$ & $n$-photon trans. \\ 
  \hline 
RRRR 
& $-3\hbar\omega \le \epsilon\le 0$ &   $n=1,2,3,4,5$\\ 
ARRR
 & $-\hbar\omega \le \epsilon\le 0$ & $1^*$  \\ 
AARR
 & $-2\hbar\omega \le \epsilon\le -\hbar\omega$ &  $n=1,2^*,3$  \\ 
AAAR
& $-3\hbar\omega \le \epsilon\le -2\hbar\omega$  &  $n=1,2,3,4,5$
 \\  
 \hline \hline
\end{tabular}}
\end{center}
\end{table}

The total nonlinear response is determined by the sum of {\em all} the channels.
As we can see in Fig. \ref{fig:PXXXX}a, both RRRR and AAAR channels show step-like transitions at the four- and five-photon resonances, i.e. at
$\hbar\omega \approx \mu_0/2 $ and $\hbar\omega \approx 2\mu_0/5$, but with opposite sign.
The sum of these two contributions would exactly cancel out in the clean limit $U=0$.
Such cancellation is however just partial for finite $U$ (or finite $\Gamma$), leaving
finite spectral structures at $\hbar\omega \approx \mu_0/2 $ and $\hbar\omega \approx 2\mu_0/5$,
as seen in  Fig. \ref{fig:PXXXX}c.
The behavior as a function of the scattering strength is shown in Fig. \ref{fig:PXXXX}d.
As mentioned above, in the clean limit $U\to 0$
the multi-photon resonances at $n=4,5$ in the individual channels
$P^{\rm RRRR}$, $P^{\rm ARRR}$, $P^{\rm AARR}$ and $P^{\rm AAAR}$
cancel out exactly and they are thus absent in the total response.
However, such cancelation is not perfect in the presence of a finite electron-impurity scattering,
leaving residual spectral structures at  frequencies corresponding to the four- and five-photon resonances.
The spectral weight of these multi-photon structures scales with the impurity scattering itself.
The absence of four- and five-photon transitions in the clean (non-interacting) limit reveal thus that these transitions
are in fact {\it incoherent-transitions} with mixed inter- and intra-band characters. 
The partial intra-band character is highlighted by the transition weight being proportional to the relaxation rate $\Gamma$,
whereas the partial inter-band character of such features is pointed out by their lying at finite frequency,
i.e. at exactly the four- and five-photon resonance energy with the peak positions not affected by of $U$.
It is worth emphasizing that these incoherent-transitions at finite frequency pinned by $\mu_0$
are a peculiar property of nonlinear optical conductivity
and they do not emerge in the linear optical conductivity.

Now, we focus on the role of the many-body
self-energy renormalization in determining the spectral features of the optical
third-harmonic generation response. 
As discussed above, $n$-photon transitions can be theoretically identified
by enforcing the condition ${\rm Re}[S(z_i)+S(z_j)]=0$
together with $\epsilon=\epsilon_{\rm min}, \epsilon_{\rm max}$,
where $\epsilon_{\rm min}$, $\epsilon_{\rm max}$ are the
lower and upper energy integration limits for each channel.
With such prescription, neglecting the many-body self-energy,
the $n$-photon transitions occur at $\hbar\omega \approx 2 \mu_0/n$,
where $\mu_0$ is the bare chemical potential.
It is however straightforward to check that the same holds true
also in the presence of (frequency-dependent) elastic scattering driven
by disorder/impurity preserving the mirror symmetry 
with respect to the Dirac point.
The analysis of linear optical conductivity is enlightening on this point.
In similar way as in the third-order response function,
at $T=0$ the edge of the interband optical transitions 
is determined by  the conditions
${\rm Re}[S(\epsilon)+S(\epsilon+\hbar\omega)]=0$ and ${\rm Re}[S(\epsilon)+S^\ast(\epsilon+\hbar\omega)]=0$, respectively,
together with the constraints $\epsilon=0$, $\epsilon=-\hbar\omega$ determined by the window of energy
integration over $\epsilon$.
We obtain thus that the edge of optical interband transitions is determined by
\begin{equation}
\hbar\omega
=
2\mu_0-\mbox{Re}\Sigma(0)-\mbox{Re}\Sigma(-2\mu_0),
\end{equation}
with a spectral weight $I_{\rm sw}$ that scales as:
\begin{equation}
I_{\rm sw}
=
\left|\mbox{Im}\Sigma^\nu(0)-\mbox{Im}\Sigma^{\nu^\prime}(-2\mu_0)\right|,
\end{equation}
where $\nu,\nu^\prime=\mbox{A,R}$. 
The elastic impurity scattering self-energy respect the following symmetry relation owing to symmetry of Dirac dispersion: 
\begin{align}
\mbox{Re}\Sigma(-\mu_0-\epsilon)
&=
-\mbox{Re}\Sigma(-\mu_0+\epsilon),
\label{reself}
\\
\mbox{Im}\Sigma^{\rm R/A}(-\mu_0-\epsilon)
&=
\mbox{Im}\Sigma^{\rm R/A}(-\mu_0+\epsilon),
\label{imrr}
\\
\mbox{Im}\Sigma^{\rm A/R}(-\mu_0-\epsilon)
&=
-\mbox{Im}\Sigma^{\rm R/A}(-\mu_0+\epsilon)
.
\label{imra}
\end{align}
These relations imply in a direct way that: ($i$) the interband optical edge
in the linear optical conductivity is determined only by $\mu_0$
and not by the renormalized chemical potential $\mu$; ($ii$) only the retarded/retarded channel
is responsible for the $n=1$ photon transition observed thus in linear optics.
Similar argumentations holds true in a straightforward way in nonlinear optics where we conclude that:
($iii$) the $n$-photon transitions are determined only by $\mu_0$
and not by the renormalized chemical potential $\mu$; ($iv$) only retarded/retarded or advanced/advanced transitions
show a sizable spectral weight and can be observed thus in the optical features.

\begin{figure}[t]
\hspace{-5mm}
\begin{overpic}[width=0.4\textwidth]{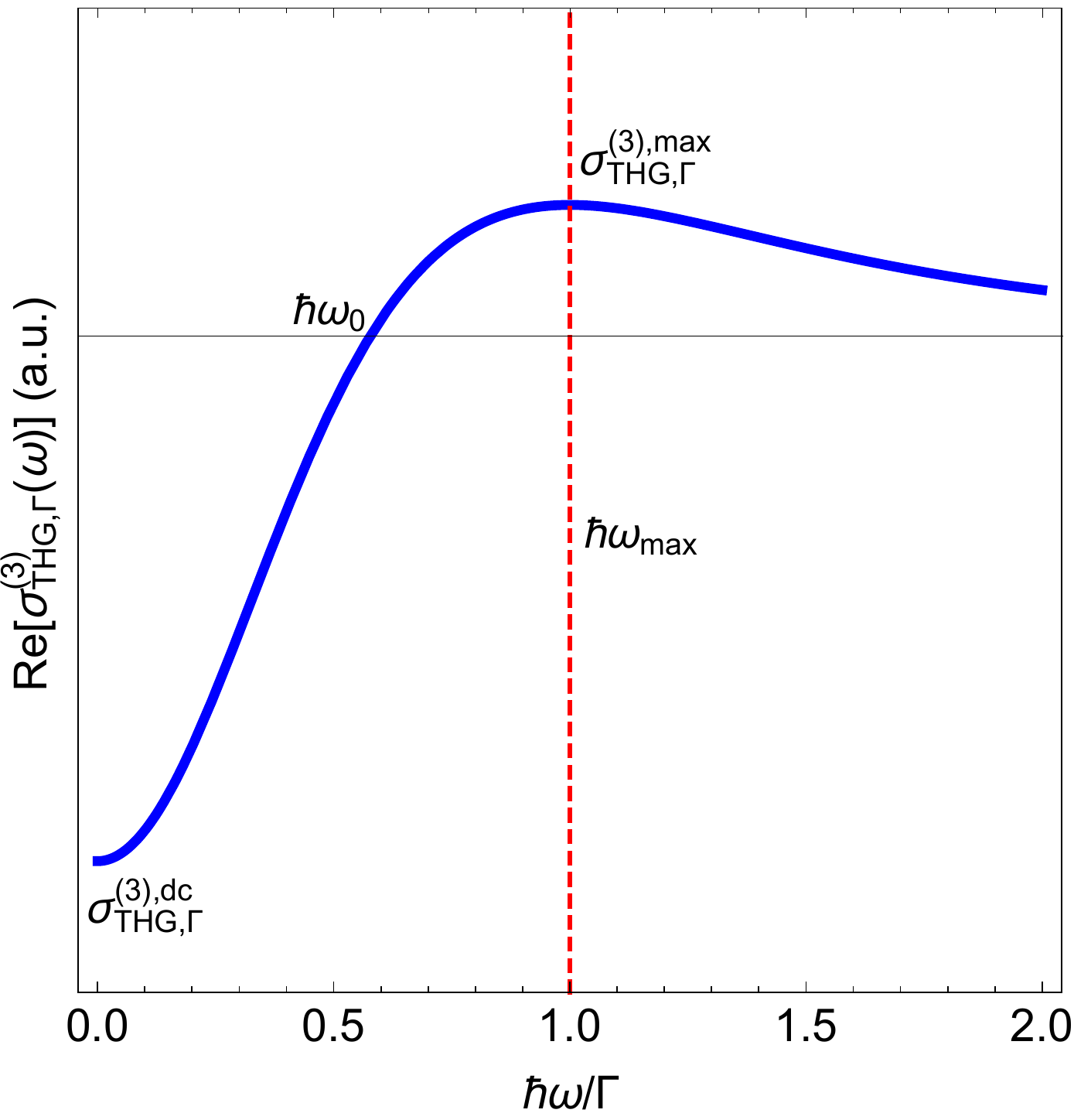} \put(25,85){}\end{overpic}
\caption{ Schematic graph of Eq.~(\ref{eqdmf}) to define parameters introduced in the text.}
\label{figSchem}
\end{figure} 
\begin{figure*}[t]
\begin{overpic}[width=1\textwidth]{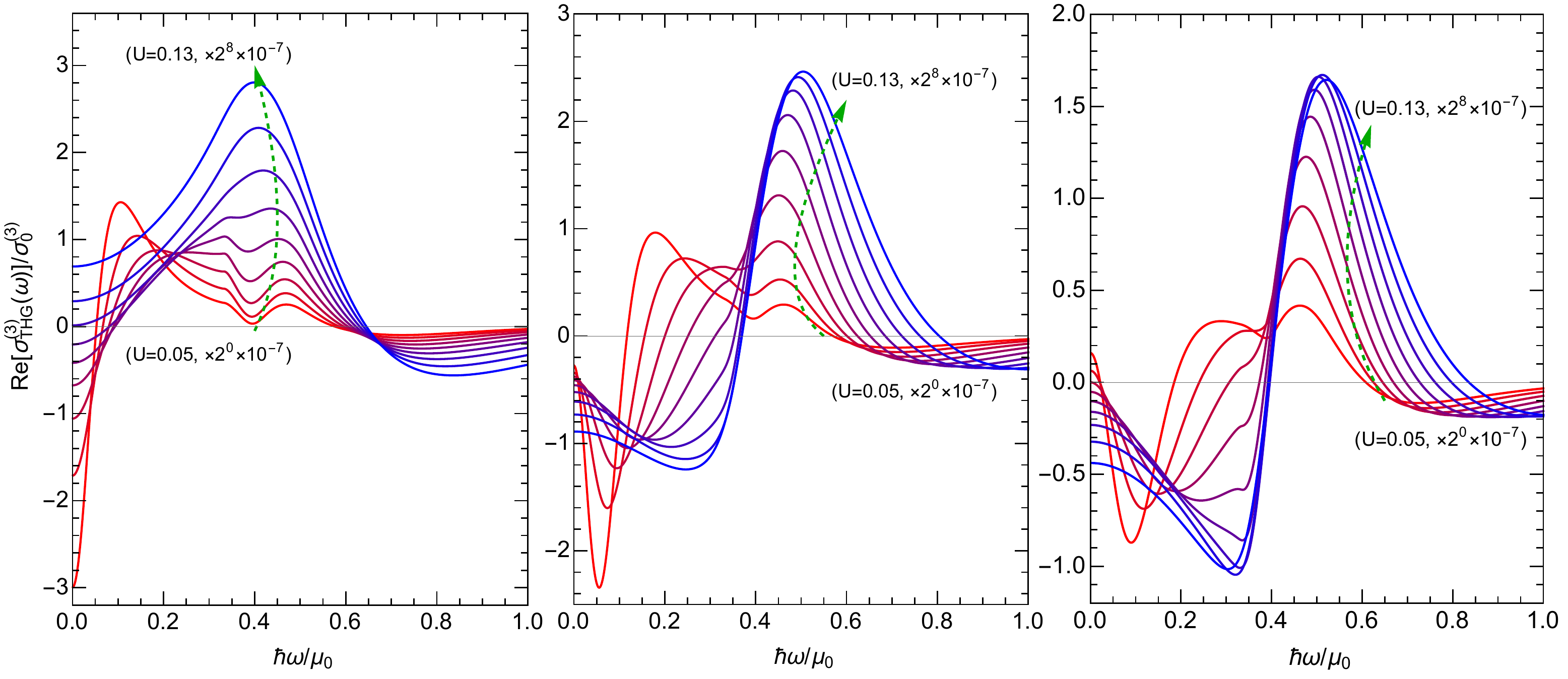} 
\put(28,39){\large (a)} 
\put(39,39){\large (b)} 
\put(72,39){\large (c)} 
\put(11,10){full quantum theory}
\put(50,10){$X_2=0$}
\put(83,10){no vertices}
\end{overpic}
\caption{(a): Real part of the intra-band THG conductivity computed at different levels of approximations:
(a) full quantum theory including all diagrams as in Fig.~\ref{fig:chi3}; (b) neglecting two-photon vertex renormalization
by setting $X_2=0$; (c) neglecting {\em all} vertex renormalization processes,
i.e. neglecting all filled vertex symbols in Fig.~\ref{fig:chi3} and retaining only the bare
one-photon term (empty circle vertex).
We set $U\in[0.05,0.06,\dots, 0.13]$ and $\mu_0=25$THz and, for sake of a better display, each curve 
is multiplied by a different magnifying factor reported in the legend. }
\label{figVertex}
\end{figure*} 

The strict validity of these symmetry relations is affected in the presence
of inelastic scattering where the even/odd symmetries with respect to the Dirac points are lifted.
Considering however the realistic case of low-energy inelastic scattering (e.g, phonons),
the breaking of the symmetry relations in the imaginary part in Eqs. (\ref{imrr})-(\ref{imra})
is limited to a narrow shell around the
Fermi level, affecting thus only the $\hbar\omega \approx 2 \mu_0$ resonance ($n=1$).
Inelastic scattering might thus reduce a bit the spectral weight of $n=1$ photon transitions
in the RRRR, AARR, AAAR channels and it might induce a finite spectral weight in the ARRR channel
which is predicted to be null under the above symmetry conditions valid however
only for elastic scattering.
Inelastic scattering would affect as well the symmetry relation for the real part of the self-energy,
as it can be obtained by Kramers-Kronig transform
of the imaginary part of the self-energy.
Since the imaginary part of the self-energy is typically affected only
in a narrow region around the Fermi level, we expect deviations
from the symmetry relation in the real part in Eq. (\ref{reself}) to be relatively small
and scaling with the strength of the inelastic coupling.
Multi-photon transitions might be just slightly shifted from the edge determined by $\mu_0$.
\subsection{Intraband THG conductivity}
The analysis of the THG optical conductivity allows to investigate in details
also the low-energy spectral features associated with the pure intraband processes,
which are not easily visible in the dimensionless function $f_3$.
A phenomenological model previously obtained \cite{Cheng_prb_2015,Mikhailov_prb_2016} based on density-matrix formalism using a constant scattering rate can qualitatively describe low-frequency profile of THG conductivity in the Boltzmann regime where $\mu\gg \Gamma,\hbar\omega$:  
\begin{equation}
\sigma^{(3)}_{{\rm THG},\Gamma}
=
 i\frac{C}{\mu(\hbar\omega+i\Gamma)^3},
 \label{eqdmf}
\end{equation}
where $C>0$ is a constant and
where  $\Gamma$ stands 
for a constant (frequency-independent) scattering rate.
Eq. (\ref{eqdmf}) captures a similar physics as the Drude term
in the linear optics.
In similar way as the Drude term, and 
unlike the interband counterpart ruled by $\mu_0$,
the spectral properties of Eq. (\ref{eqdmf})
are governed only by $\Gamma$ and by the ratio $\hbar \omega/\Gamma$. 
Intraband transitions occur at the Fermi surface and accordingly the intraband THG conductivity depends on the renormalized chemical potential $\mu$. 
Remarkably, the above model predicts for any strength of $\Gamma$
a negative dc conductivity $\sigma^{(3),{\rm dc}}_{\rm THG,\Gamma} = -C/\mu\Gamma^3<0$ in the limit $\omega\to0$,
as well as a maximum $\sigma^{(3),{\rm max}}_{{\rm THG},\Gamma}=C/4\mu\Gamma^3$ at $\hbar\omega_{\rm max} = \Gamma$ and a zero $\sigma^{(3)}_{{\rm THG},\Gamma} (\omega=\omega_0)=0$ with $\hbar\omega_0 = \Gamma/\sqrt{3}$.  
See Fig.~\ref{figSchem} for the graphical clarification of parameters defined above. 
\begin{figure*}[t]
\hspace{-5mm}
\begin{overpic}[width=0.40\textwidth]{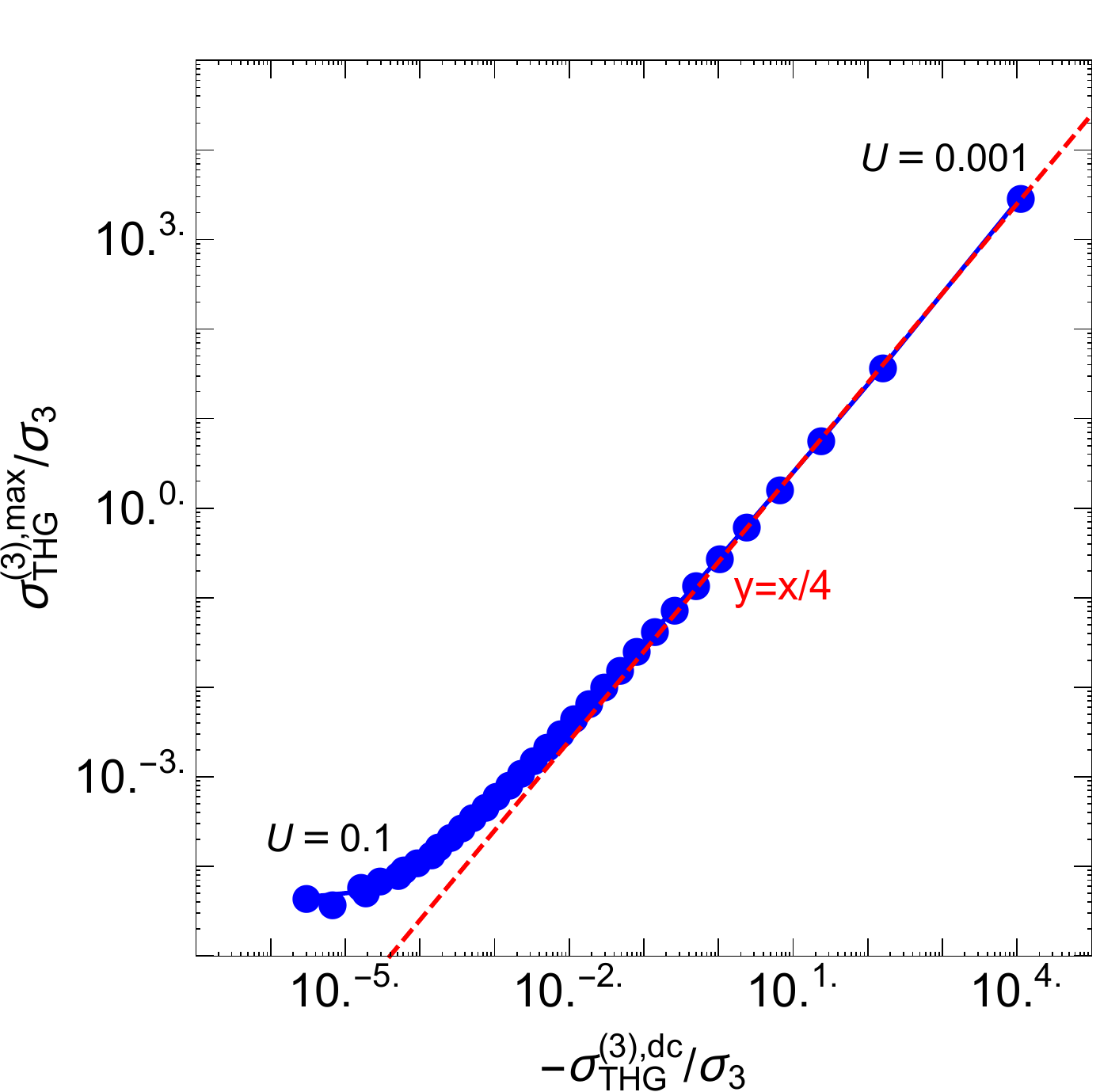} \put(25,85){\large (a)}\end{overpic}
\hspace{3mm}
\begin{overpic}[width=0.40\textwidth]{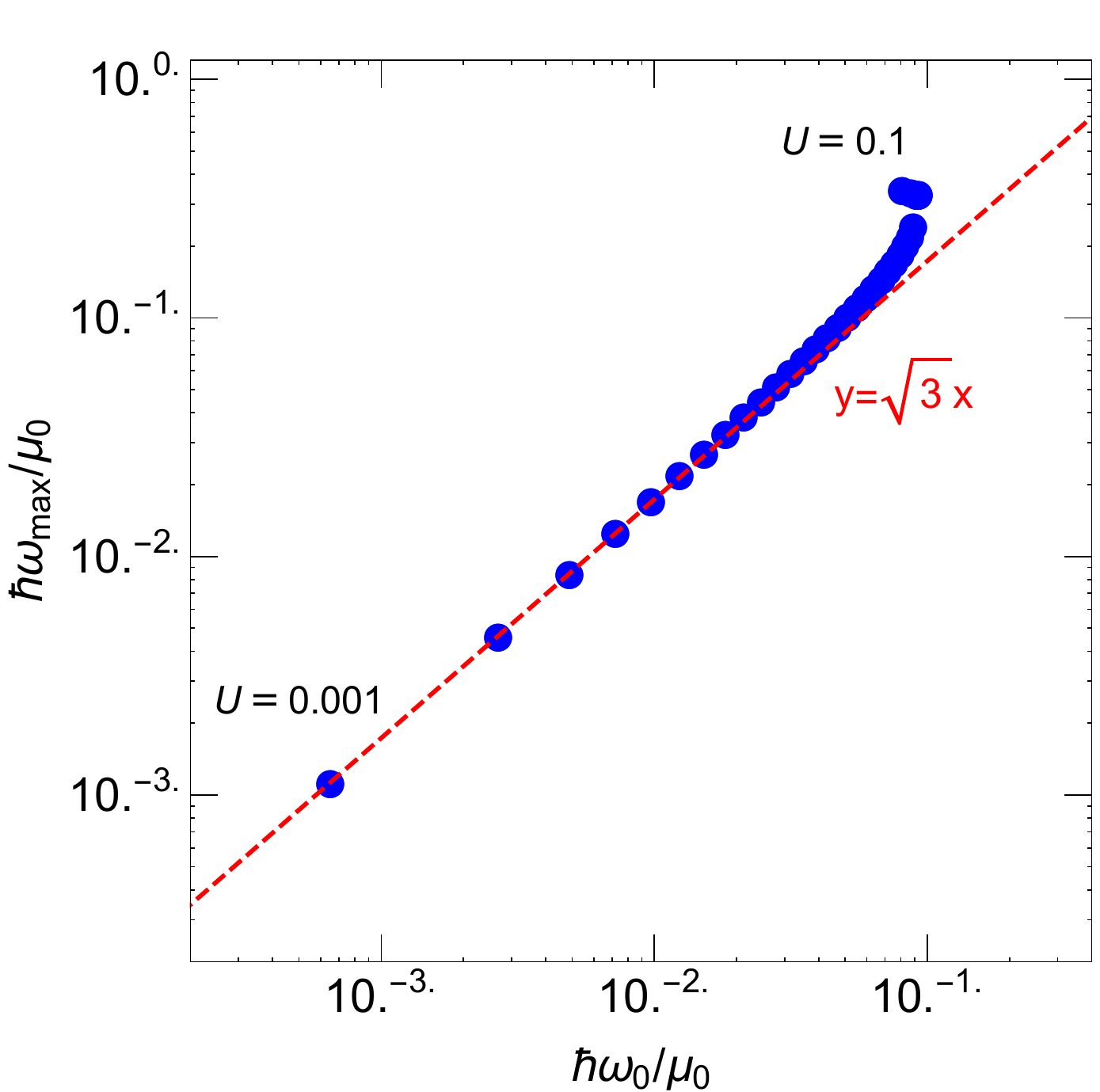} \put(25,85){\large (b)}\end{overpic}
\begin{overpic}[width=0.42\textwidth]{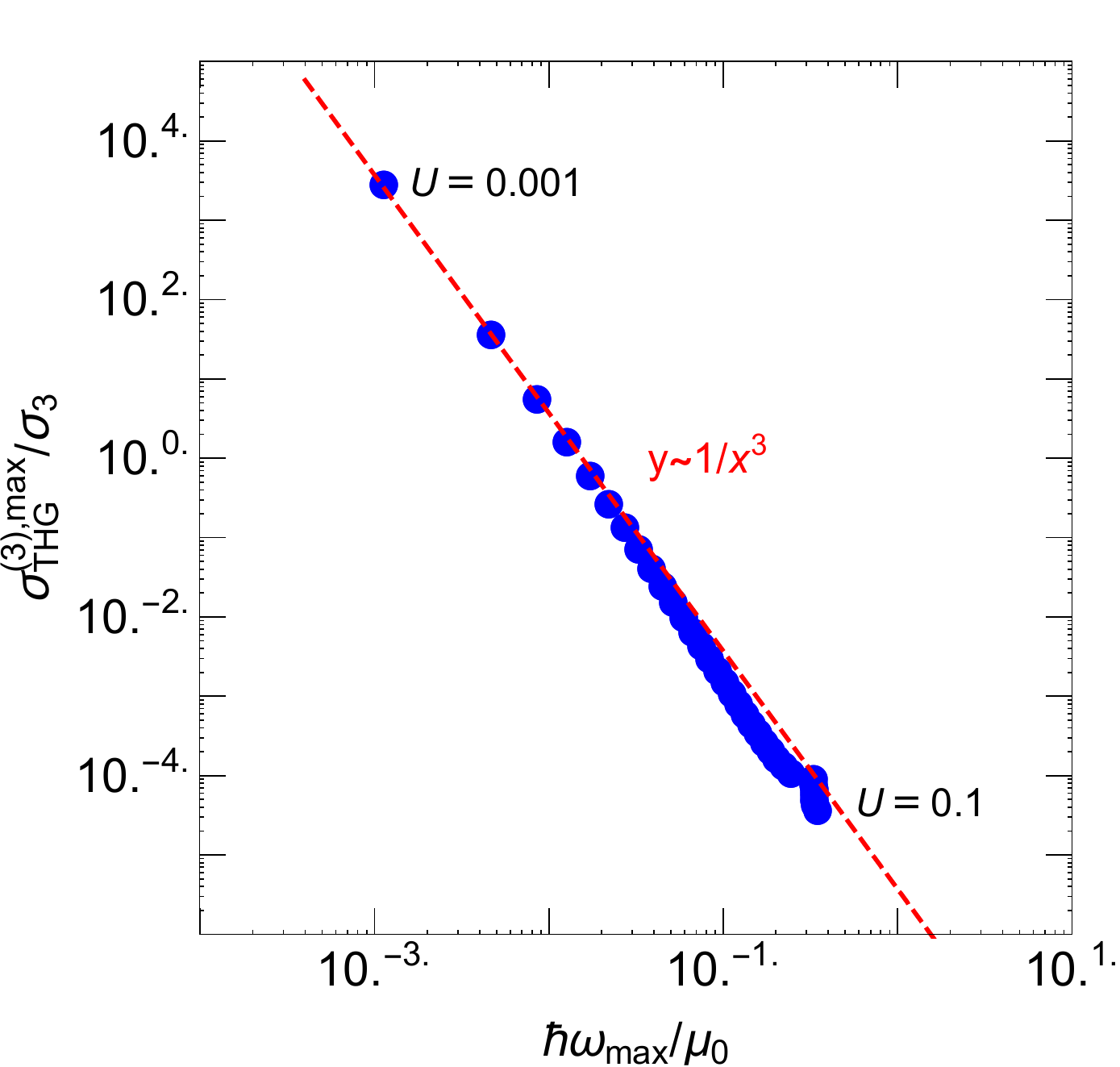} \put(80,80){\large (c)}\end{overpic}
\begin{overpic}[width=0.42\textwidth]{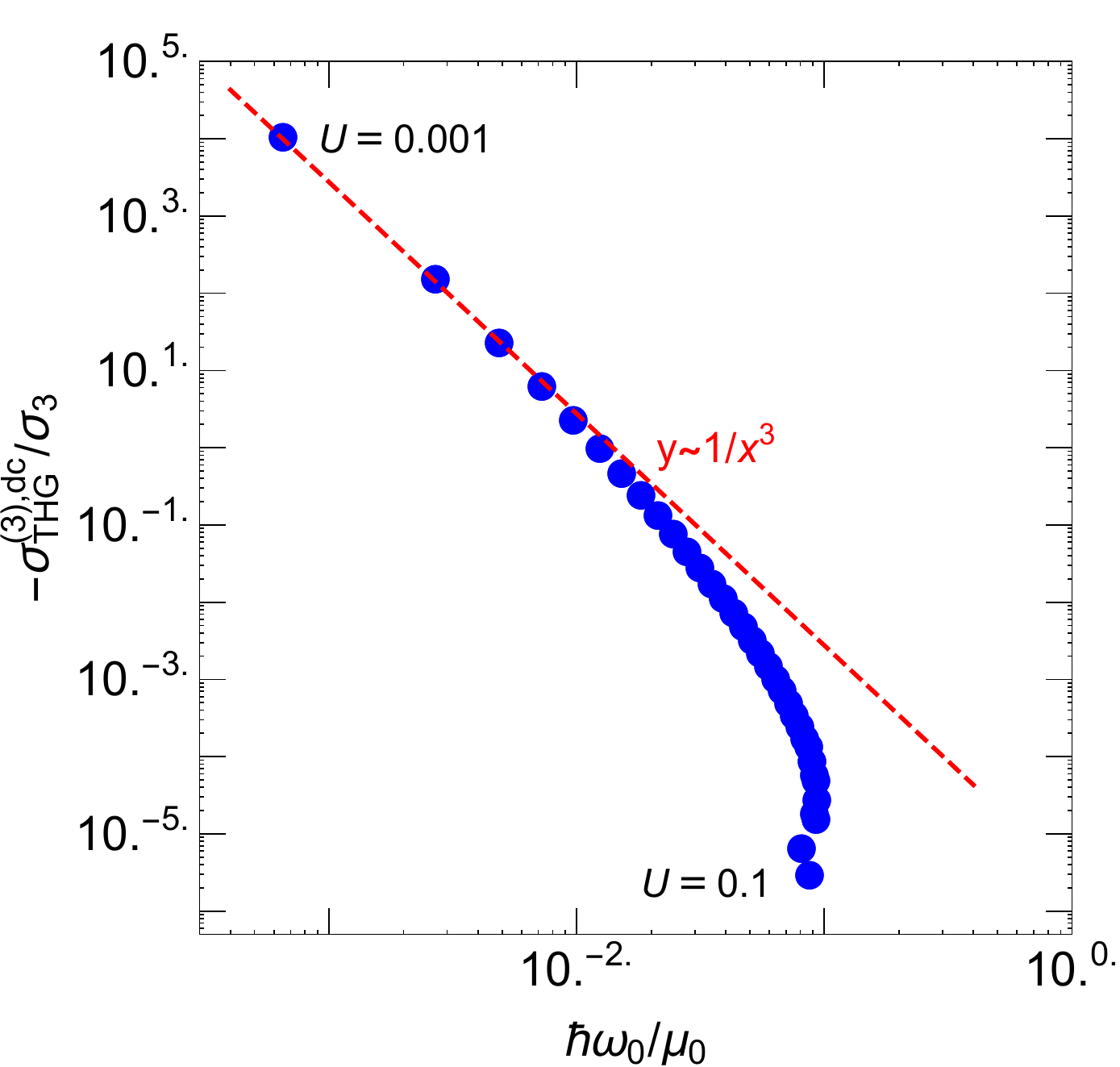} \put(80,80){\large (d)}\end{overpic}
\caption{Comparison between full many-body theory with the phenomenological model given in  Eq.~(\ref{eqdmf}). (a) Log-Log scale plot for the maximum value of real part of the intra-band THG conductivity versus the third-order dc conductivity compared with linear scaling dashed red line obtained from the phenomenological formula.  
(b) Log-Log scale plot for $\omega_{\rm max}$ versus $\omega_0$ which is compared with linear scaling line (dashed red line) based on the phenomenological model. 
(c) Cubic scaling of the maximum value of real part of the intra-band THG versus $\omega_{\rm max}$ which is valid only for small values of $U$. 
(d) Cubic scaling for third-order dc conductivity versus $\omega_0$ which is valid only for small values of $U$. 
Note that $\sigma_3= 10^{10} \sigma^{(3)}_0$ and we set $\mu_0=25$THz for this figure. }
\label{figPhenom}
\end{figure*} 

In Fig. \ref{figVertex}a we show the real part of the THG optical conductivity
as obtained by our quantum conserving approach
for different impurity scattering strengths.
In the weak scattering limit (red curves), the four- and five-photon edges are 
visible in the bump and in the dip at $\hbar \omega=0.5\mu_0$ and $\hbar \omega=0.4\mu_0$, respectively.
Besides these spectral structures, with mixed intra-interband character,
we are able now to reveal a purely intraband low frequency structure with a strong resemblance
with the phenomenological model of $\sigma^{(3)}_{{\rm THG},\Gamma}$ given in  Eq. (\ref{eqdmf}).

Giving the complex nature of the nonlinear response, even in the intraband dc limit
where the multi-photon vertex renormalization has shown to be highly relevant,\cite{Rostami_Cappeluti_arxiv_2020}
the robustness of the phenomenological constant-$\Gamma$ model
appears by no way trivial.
This issue is pointed out in Fig. \ref{figVertex}b,c where we plot the THG optical conductivity
at two levels of (non-conserving) approximations:
neglecting the two-photon vertex renormalization ($X_2=0$), that has
been shown to be dominant in the dc limit, see panel (b); and
neglecting {\em all} the vertex renormalization processes, see panel (c).
For both cases we find a completely different scenario at low-frequency
with respect to the phenomenological constant-$\Gamma$ model
and to the compelling numerical results.
We remind that the constant-$\Gamma$ model is itself a {\em conserving} approximation,
where to a field-independent and frequency-independent self-energy correspond
unrenormalized vertices.
We rationalize the agreement at weak scattering between the constant-$\Gamma$ model
and the fully many-body theory on the basis of the requirement of a conserving analysis.
Quantum many-body theories, when based on non-conserving approaches, 
might severely give spurious results.

For stronger scattering strength $U$ (see blue curves in Fig.~\ref{figVertex}) we enter to the quantum regime where the scattering rate $\Gamma$ is comparable (or larger) than the chemical potential $\mu$. For this case, we observe a full merging of all spectral features (pure intraband peak merges with four and five-photon incoherent structures) and morphing into a single peak as well as the nonlinear dc conductivity changes sign. This sign change is a result of two-photon vertex self-generation which was discussed previously \cite{Rostami_Cappeluti_arxiv_2020}. Obviously, the phenomenological model given in Eq.~(\ref{eqdmf}) is no longer valid in this strong interacting regime.

In order to investigate in a deeper way the comparison of our many-body theory with the phenomenological model in the weak scattering limit, we analyze the numerical output
of the fully quantum conserving theory in the low-frequency range within
the ansatz of the phenomenological model of Eq. (\ref{eqdmf}).
More precisely, we determine in the full quantum theory, as a function of the scattering strength $U$,
the dc limit $\sigma^{(3),{\rm dc}}_{\rm THG}=\sigma^{(3)}_{\rm THG}(\omega=0)$;
the frequency $\omega_0$ where the THG has the first low-frequency zero;
the frequency $\omega_{\rm max}$ where the THG conductivity has its first maximum;
the value $\sigma^{(3),{\rm max}}_{\rm THG}$ of the THG conductivity at $\omega=\omega_{\rm max}$.
In Fig. \ref{figPhenom}a-d we compare
the mutual dependence of the such characteristic spectral parameters 
as obtained from the numerical results and as estimated from
the model in Eq. (\ref{eqdmf}).
We find an excellent agreement in the very weak scattering regime,
but a quick deviation by increasing the scattering strength $U$. 
Keeping in mind that in the phenomenological model $\hbar \omega_{\rm max}\approx \Gamma$,
we estimate a breakdown of the constant-$\Gamma$ model for $\Gamma \approx 0.01 \mu$, e.g. see Fig.~\ref{figPhenom}d. On the basis of Fig. \ref{figVertex}a, we rationalize this  failure as the overlap
of the purely intraband term with the five- and four-photon edge transitions,
starting from $\hbar \omega \approx 0.4\mu_0-0.5 \mu_0$.

\section{Summary and Conclusion}\label{sec:summery_conclusion}
In this paper we have explored the effects
of elastic impurity scattering effect on the third-harmonic generation of graphene by using a conserving diagrammatic method
that includes self-consistently both self-energy and vertex renormalization contributions.  
As a result of the field dependence of the self-energy, we showed that
interaction-induced multi-photon vertex diagrams are relevant.
In particular we have predicted the onset of novel incoherent resonances at four and five-photon transition energy
with mixed intra- and inter-band character.
Furthermore, we have shown that the main features of the purely intraband contribution in the full quantum theory
are qualitatively comparable with phenomenological models that assume a frequency/field independent self-energy,
whereas non-conserving approaches give rise to spurious results.
In the terahertz regime the proximity of the five-photon transition edge at $\hbar \omega=2\mu_0/5$
might affect the interpretation of experimental data in terms of a purely intraband term.

Our study, shed light on the importance of many-body interaction on the qualitative explanations of nonlinear optics in the terahertz and infrared (intra-band) regimes. Although the numerical study is performed for the third-harmonic generation, the formalism is quite general and applicable for two-photon absorption, nonlinear Kerr effect, etc. Moreover, our study can be simply generalized to explain terahertz nonlinear response in other two-dimensional materials such as transition-metal dichalcogenides and twisted bilayer graphene. 

\section{Acknowledgements}
H.R. acknowledges the support from the Swedish Research Council (VR 2018-04252).
\bibliography{bibliography}
\newpage
\appendix 

 \section{Renormalization of the one-photon vertex}\label{app:vertex_1}
The one-photon vertex renormalization is depicted diagrammatically in Fig.~1b of the main text and it reads
\begin{align}
 &\hat\Lambda_1({\bf p},{\bf p}+{\bf q};n,n+m)
 = \hat\lambda_1({\bf p},{\bf p}+{\bf q};n,n+m) 
\nonumber\\&+ \gamma_{\rm imp}  
\sum_{{\bf k}} \hat G({\bf k},n)  \hat \Lambda_1({\bf k},{\bf k}+{\bf q};n,n+m) \hat G({\bf k}+{\bf q},n+m)~,
\end{align}
where $m\equiv iq_m$ and $n \equiv ik_n=ip_n$ stand for the bosonic and fermionic Matsubara frequencies, respectively. Note that in the integrand we have shifted the dummy momentum $\bf k$ as ${\bf k} +{\bf p}\to {\bf k}$ and therefore we can  see that vertex correction does not depends on the fermion momentum $\bf p$. 
For the optical (or dipole) approximation we have $\bf q=0$.  Therefore, the Bethe-Salpeter relation for the one-photon vertex function reads   
\begin{align}
&\hat\Lambda_{1}(n,n+m)= \hat \lambda_1(n,n+m)
\nonumber\\&
+ \gamma_{\rm imp} \sum_{{\bf k}}
 \hat G(k,n)\hat\Lambda_{1}(n,n+m) \hat G (k,n+m)~.
\end{align}
Note that we have $\hat \lambda_1(n,n+m)= \hat \Lambda^{(0)}_1 ={\delta \hat G^{-1}_0}/{\delta A}|_{\bf A\to 0} = -ev \hat\sigma_y$ where $\hat G_0$ stands for the non-interacting Green's function. 
We assume the following ansatz for the vertex function 
\begin{align}
\hat \Lambda_1(n,n+m) = a\hat I+b\sigma_x+(c+v)\sigma_y+d\sigma_z~.
\end{align}
Using the fact that the integral of odd-function of ${\bf k}$ is zero, we obtain $a=b=d=0$ and eventually the following result for the vertex function $\hat \Lambda_1 =  (-ev \hat\sigma_y) \Lambda_1$ and $\hat \lambda_1 =  (-ev \hat\sigma_y) \lambda_1$ where
\begin{align}
 \Lambda_1(n,n+m)  =   Q_1(n,n+m) \lambda_1 ~. 
 \label{eql1}
\end{align}
Note that $\lambda_1=1$ and we defines the one-photon vertex renormalization factor:
\begin{align}
 Q_1(z,z')  = \frac{1}{1- U X_1(z,z') }~. 
\end{align}
in whcih
\begin{align}
 X_1(z,z') = \frac{\gamma_{\rm imp}}{2 U} \sum_{\bf k} {\rm Tr}[\hat\sigma_y \hat G({\bf k},z)\hat\sigma_y \hat G({\bf k},z')]~.
\end{align}
Using dimensional regularization, we find the following formula for the $X_1$ function: 
\begin{align}
X_1(z,z')  =  \frac{S(z)S(z')}{S(z)^2-S(z')^2}\ln\left[\frac{S(z')^2}{S(z)^2}\right]~.
\end{align}
Obviously $X_1(z,z') = X_1(z',z)$. 
\section{Renormalization of the two-photon vertex}\label{app:vertex_2}
Similar to the one-photon vertex case, it can be shown that the two-photon vertex function is independent of the fermionic momentum, $\bf p$. Moreover, for the optical limit we can neglect the photon momentum $\bf q$. 
The self-consistent Bethe-Salpeter relation for the two-photon vertex function is depicted in Fig.~1c of the main text and it reads 
\begin{align}\label{eq:BS_2photon}
&\hat \Lambda_{2}(n,n+m,n+2m)  = 
\hat {\lambda}_{2} (n,n+m,n+2m)
\nonumber\\&+ \gamma_{\rm imp}  
\sum_{\bf k} \hat G({\bf k},n)  \hat \Lambda_{2} (n,n+m,n+2m) \hat G({\bf k},n+2m)~. 
\end{align}
In the non-interacting Dirac system the ``bare''    
two-photon vertex function is zero, $\hat \Lambda^{(0)}_{2} \propto \delta^2 \hat G^{-1}_0/\delta A^2|_{\bf A\to 0}=0$, due to the linear momentum dependence of the Hamiltonian. However, due to interaction the unrenormalized two-photon vertex $\hat \lambda_{2}$ is finite given by the following relation (see Fig.~1e of the main text) 
\begin{align}
&\hat {\lambda}_{2} (n,n+m,n+2m) = 
-\gamma_{\rm imp}  
\sum_{{\bf k}} \hat G({\bf k},n) 
\hat \Lambda_y(n,n+m) 
 \nonumber\\&\times 
\hat G({\bf k},n+m)
\hat \Lambda_y(n+m,n+2m)  
\hat G({\bf k},n+2m)~. 
\end{align} 
From now on we adopt the short-hand notation $z_j = n+j m$ with $j=0,1,2,3$. 
We find $\hat \lambda_{2} =  (-ev\hat\sigma_y)^2  \lambda_2$ with $\hat\sigma^2_y=\hat I$
and 
\begin{align}
\lambda_2(z_0,z_1,z_2) =  
Q_1(z_0,z_1)Q_1(z_1,z_2)U   Z(z_0,z_1,z_2)~,
\end{align} 
in which $Q_1(z_i,z_j)$ is the one-photon renormalization factor defined in the previous subsection,
and where
\begin{align}
 Z(z_0,z_1,z_2) = -\frac{\gamma_{\rm imp }}{2 U} \sum_{{\bf k}} {\rm Tr}[\hat G({\bf k},z_0) \hat \sigma_y 
\hat G({\bf k},z_1)
\hat \sigma_y  \hat G({\bf k},z_2)]~.
\end{align}
By performing the momentum integration using the dimensional regularization, we obtain 
\begin{align}
Z(z_0,z_1,z_2)  = \frac{X_1(z_0,z_1) - X_1(z_1,z_2)}{S(z_0)-S(z_2)}~.
\end{align}
By solving the the self-consistent Bethe-Salpeter relation for the two-photon vertex given in Eq.~(\ref{eq:BS_2photon}), we obtain $\hat \Lambda_{2}= (-ev\hat\sigma_y)^2 \Lambda_2$  with   
\begin{align}
  \Lambda_{2}(z_0,z_1,z_2)  = 
  Q_2(z_0,z_2) 
  \lambda_{2}(z_0,z_1,z_2)
  ,
  \end{align}
  in which $Q_2(z_0,z_2)$ is the two-photon Bethe-Salpeter renormalization factor
 \begin{align}
 Q_2(z_0,z_2) 
 =
 \frac{1}
{1-U X_2(z_0,z_2)},
\end{align}
 and
where  
\begin{align}
X_2(z,z') = \frac{\gamma_{\rm imp}}{2 U} \sum_{\bf k} {\rm Tr}[\hat G({\bf k},z) \hat G({\bf k},z')].
\end{align}
Therefore, the two-photon renormalization factor reads  
\begin{align}
Q_2(z,z') =  \frac{S(z)-S(z')}{z-z'}~.
\end{align}
\section{Renormalization of the three-photon vertex}\label{app:vertex_3}
Similar to the case of two-photon case, the impurity scattering induces a finite three-photon vertex as defined in Fig.~1f of the main text.
Accordingly we find  $\hat{\lambda}_{3}= (-ev\hat \sigma_y)^3  \lambda_3$ with 
\begin{align}
\lambda_3(z_0,z_1,z_2,z_3)  =   \sum^3_{n=1}M_n(z_0,z_1,z_2,z_3)~,
\end{align}
where
\begin{align}
 M_1(z_0,z_1,z_2,z_3) 
& =
 U \Omega_1(z_0,z_1,z_2,z_3)Q_1(z_0,z_1)
  \nonumber\\&\times  Q_1(z_1,z_2)Q_1(z_2,z_3)
 ,
\\
 M_2(z_0,z_1,z_2,z_3)  
 &=
U  \Omega_2(z_0,z_2,z_3) \lambda_2 (z_0,z_1,z_2)  \nonumber\\&\times Q_1(z_2,z_3)Q_2(z_0,z_2)
,
\\
 M_3(z_0,z_1,z_2,z_3) 
 &=
U \Omega_3(z_0,z_1,z_3) \lambda_2(z_1,z_2,z_3) \nonumber\\&\times Q_1(z_0,z_1)Q_2(z_1,z_3)
.
 \end{align}
Here $Q_1(z_i,z_j)$, $Q_2(z_i,z_j)$ are the Bethe-Salpeter one- and two-photon renormalization functions, 
respectively.
The explicit expression for  $\Omega_1$ function is given by   
\begin{align}
 \Omega_1(z_0,z_1,z_2,z_3)
 = \sum^3_{n=1} u_n(z_0,z_1,z_2,z_3)  \ln\left[\frac{S(z_0)^2}{S(z_n)^2}\right]~,
\end{align}
where we have 
\begin{align}\label{eq:un}
u_n(z_0,z_1,z_2,z_3)= \frac{\Pi_{j}S(z_j)+ R(z_0,z_1,z_2,z_3) S(z_n)^2}
{\Pi_{j \neq n}[S(z_n)^2-S(z_j)^2]}~,
\end{align}
in which $R(z_0,z_1,z_2,z_3)=S(z_0)S(z_2)+S(z_1)S(z_3)$. 
Similarly, by explicit calculation of the momentum integration, one can obtain  
$\Omega_2(z_0,z_2,z_3) = Z(z_2,z_3,z_0)$
and
$\Omega_3(z_0,z_1,z_3) = Z(z_3,z_0,z_1)$. 
Finally, the Bethe-Salpeter renormalization of the three-photon vertex function gives:
\begin{align}
 \Lambda_{3}(z_0,z_1,z_2,z_3) = Q_3(z_0,z_3) \lambda_{3}(z_0,z_1,z_2,z_3)~.
\end{align}
where  $Q_3 (z_1,z_2) = Q_1 (z_1,z_2)$. 
\end{document}